\documentclass[Journal]{IEEEtran}
\def\doublecolumn{1}

\usepackage{amsmath, amsfonts, amssymb, bbm}
\usepackage{graphicx,dblfloatfix}
\usepackage[space]{cite}
\usepackage{hyperref}
\usepackage{color}
 
\newtheorem{theorem}{Theorem}
\newtheorem{corollary}{Corollary}[theorem]
\newtheorem{lemma}{Lemma}
\newtheorem{proposition}{Proposition}
\newtheorem{definition}{Definition}

\newtheorem{remark}{Remark}
\newtheorem{example}{Example}

\allowdisplaybreaks[4] 
\usepackage{color}

\newcounter{tempEquationCounter} 
\newcounter{thisEquationNumber}
\newenvironment{floatEq}
{\setcounter{thisEquationNumber}{\value{equation}}\addtocounter{equation}{1}
\begin{figure*}[!t]
\normalsize\setcounter{tempEquationCounter}{\value{equation}}
\setcounter{equation}{\value{thisEquationNumber}}
}
{\setcounter{equation}{\value{tempEquationCounter}}
\hrulefill\vspace*{4pt}
\end{figure*}
}

\newcommand{\set}{{\mathcal{S}}}
\newcommand{\setK}{{\mathcal{K}}}
\newcommand{\setW}{{\mathcal{W}}}
\newcommand{\setT}{{\mathcal{T}}}
\newcommand{\setM}{{\mathcal{M}}}
\newcommand{\setC}{{\mathcal{S}^\text{c}}}

\newcommand{\R}{\mathcal{R}}
\newcommand{\bm}[1]{\mathbf{#1}}
\renewcommand{\b}[1]{\mathbb{#1}}
\renewcommand{\Pr}{\b{P}}
\newcommand{\Ps}{\mathcal{P}^*}
\newcommand{\Pe}{\mathsf{P}_\text{e}}
\newcommand{\h}[1]{\hat{#1}}

\newcommand{\ellc}{{\ell^{\text{c}}}}
\newcommand{\step}[2]{\stackrel{\textnormal{#1}}{#2}}

\newcommand{\indicator}[1]{\mathbbm{1}{\left\{ {#1} \right\} }}

\begin{document}

\title{A Multiway Relay Channel with Balanced Sources}
\author{Lawrence Ong and Roy Timo
\thanks{L.\ Ong is with the University of Newcastle (lawrence.ong@cantab.net), and R.\ Timo is with the Technische Universit\"{a}t M\"{u}nchen (roy.timo@ieee.org). This work was presented in part at the 2012 IEEE International Symposium on Information Theory, and it was supported by the Australian Research Council grant FT140100219 and the Alexander von Humboldt Foundation.}
}

\maketitle

\begin{abstract}
We consider a joint source-channel coding problem on a finite-field multiway relay channel, and we give closed-form lower and upper bounds on the optimal source-channel rate. These bounds are shown to be tight for all discrete memoryless sources in a certain class $\Ps$, and we demonstrate that strict source-channel separation is optimal within this class.  We show how to test whether a given source belongs to $\Ps$, we give a balanced-information regularity condition for $\Ps$, and we express $\Ps$ in terms of conditional multiple-mutual informations. Finally, we show that $\Ps$ is useful for a centralised storage problem.
\end{abstract}


\section{Introduction}

\IEEEPARstart{T}{he} \emph{multiway relay channel} is a multicast network model in which many users exchange data via a relay~\cite{gunduzgoldsmithpoor-13-it,ongmjohnsonit11}. The model is widely applicable to wireless cellular~\cite{andrewschoihealth07}, satellite~\cite{parkohpark09}, mesh~\cite{kattirahulhu08} networks, and storage networks~\cite{jepson04}, and its information-theoretic limits will provide design insights for future cooperative communications systems. Despite much recent attention~\cite{ongmjohnsonit11,timolechnerongjohnson12,ongjohnsonisit12,Barros-Jan-2006-A,Cui-Oct-2009-A,Cui-2012-A,rankovwittneben06,sugamal10,Wyner-Jun-2002-A}, the channel's information-theoretic limits remain largely unknown. In this work, we consider the limits of the following setup:
\begin{itemize}
\item $L \geq 2$ users have correlated data that need to be exchanged via the relay. The correlated data are generated by an arbitrary discrete memoryless source.
\item The uplink channel (users to relay) and the downlink channel (relay to users) are memoryless additive-noise channels defined over an arbitrary finite field.
\end{itemize}

The discrete memoryless source serves as a simple model for distributed correlated data in, for example, cloud storage systems, sensor networks, and mobile applications~\cite{Varshney-May-2011-A,Pradhan-Mar-2003-A,Pradhan-Mar-2002-A,Wang-April-2015-A}. The \emph{finite-field channel} both generalises the binary-symmetric channel and serves as a stepping stone to other important linear additive-noise channels, such as the Gaussian multiway relay channel.  

An efficient communications system for the above problem needs to effectively integrate distributed data compression with multiuser channel coding. For such systems, an important information-theoretic benchmark is the \emph{optimal source-channel rate}---the minimum number of channel uses per source symbol needed for reliable communications.  The main problem of interest in this paper is to determine the optimal source-channel rate. 

Ong et al.~\cite{onglechnerjohnsonkellett13} studied a limited version of the above problem with three users. They determined the optimal source-channel rate for sources with specific entropic structures, and demonstrated that strict source-channel separation is optimal for such class of sources. The present paper strengthens and generalises the main ideas and results of Ong et al.~\cite{onglechnerjohnsonkellett13} to three or more users.

In Section~\ref{Sec:MWRC}, we present lower and upper bounds on the optimal source-channel rate that hold for any source and $L \geq 2$ users. The upper bound (i.e., achievability) is proved using a standalone distributed source code proposed by Timo et al.~\cite{timolechnerongjohnson12} together with a standalone \emph{functional-decode-forward} channel code by Ong et al.~\cite{ongmjohnsonit11}. 

We show in Section~\ref{Sec:Ps} that the above lower and upper bounds coincide for a class of sources $\Ps$---regardless of the channel parameters---and the result is a closed-form expression for the optimal source-channel rate. The class $\Ps$ is computable in the usual information-theoretic sense, and it is determined by the underlying distributed source-coding problem. We show how to test whether or not any given source belongs to $\Ps$ by solving a certain linear system. 

In Section~\ref{Sec:Balanced}, we give a \emph{balanced information} regularity condition for $\Ps$ that can be used whenever the methods in Section~\ref{Sec:Ps} are either impractical or undesirable. The balanced-information condition is expressed in term of \textit{conditional multiple-mutual informations}~\cite{mcgill54,han80b,hekstrawillems89}, which can be visualised using information diagrams and the $I$-measure formalism of Yeung~\cite{yeung08}. We use this approach to determine the optimal source-channel rate of some sources. 

Finally, in Section~\ref{Sec:Storage}, we conclude the paper by considering a centralised storage problem with $L$ clients. The class of sources $\Ps$ plays an important role in this problem, and we show how the results of Sections~\ref{Sec:Ps} and~\ref{Sec:Balanced} can be used to describe the optimal storage rate. 


\section{Optimal Source-Channel Rate}\label{Sec:MWRC}


\subsection{Notation}

We denote random variables by uppercase letters, e.g. $W$; their alphabets by matching calligraphic font, e.g. $\mathcal{W}$; and elements of an alphabet by lowercase letters, e.g. $w \in \mathcal{W}$. The Cartesian product of $\mathcal{W}$ and $\mathcal{W}'$ is $\mathcal{W} \times \mathcal{W}'$, and the $m$-fold Cartesian product of $\mathcal{W}$ is $\mathcal{W}^m$. For integers $a$ and $b$, with $a \leq b$, we let $[a,b] := \{a,a+1,\ldots,b\}$. 
Subsets  and strict subsets are identified by $\subseteq$ and $\subset$ respectively. We will often consider subsets $\set \subseteq [1,L]$ and, in such cases, we let $\set^\text{c} := [1,L] \backslash \set$ denote the complement of $\set$. When $\set$ is a singleton $\{\ell\}$ or the complement of a singleton $\{\ell\}^\text{c}$, we write $\ell = \{\ell\}$ and $\ellc = \{\ell\}^\text{c}$. We let $\|\bm{r}\| :=|r_1| + |r_2| + \cdots + |r_L|$ denote the $L^1$ norm of a real-valued vector $\bm{r} \in \mathbb{R}^L$. The base of all logarithms in this paper is two.


\subsection{Source model}\label{Sec:MWRC:Source}

Consider $L$ arbitrarily-dependent discrete random variables 
\begin{equation}\label{Eqn:Source}
(W_1,W_2,\ldots,W_L), 
\end{equation}
where the $\ell$-th variable $W_\ell$ is defined on an alphabet $\setW_\ell$ and associated with user $\ell$. Let
\begin{equation*}
(\bm{W}_1,\bm{W}_2,\ldots,\bm{W}_L) := \big\{(W_{1,t},W_{2,t},\ldots,W_{L,t})\big\}_{t = 1}^m
\end{equation*}
be a string of $m$ independent and identically distributed (iid) copies of~\eqref{Eqn:Source} indexed by~$t$. The source data of user $\ell$ is the iid $m$-tuple $\bm{W}_\ell = (W_{\ell,1},W_{\ell,2},\ldots,W_{\ell,m})$. Each user is required to exchange its source data with that of every other user. 


\subsection{Channel model}\label{Sec:MWRC:Channel}

\begin{figure}[t!]
\begin{center}
\includegraphics[width=0.48\textwidth]{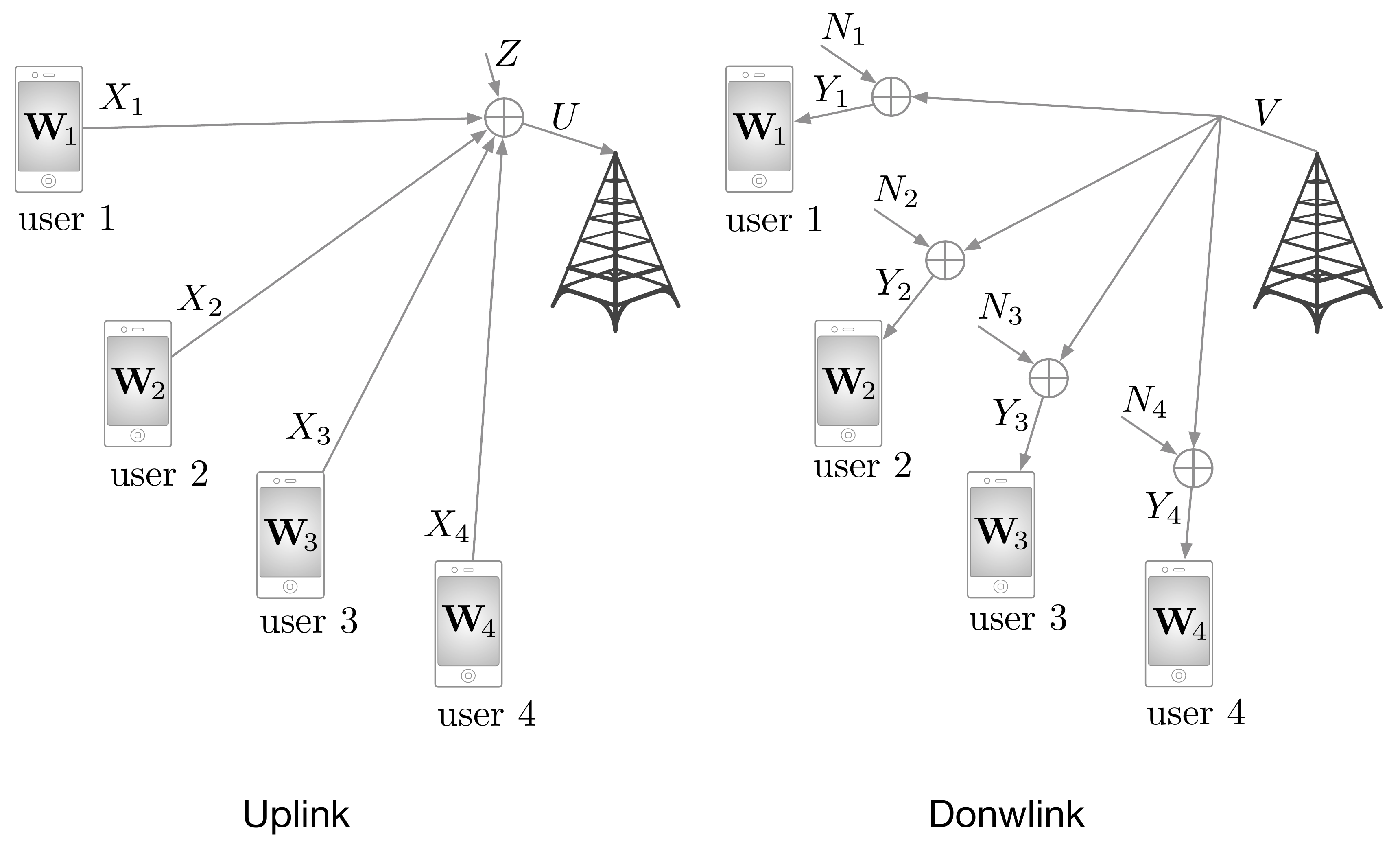}
\end{center}
\vspace{-5mm}
\caption{The uplink and downlink channel laws of the finite-field multiway relay channel with four users.}
\vspace{-5mm}
\label{Fig:MWRC}
\end{figure}

The uplink channel (users to relay) and downlink channel (relay to users) are both memoryless and defined over a finite field $\mathcal{F}$ equipped with addition $\oplus$. The per-symbol law characterising the memoryless uplink channel is
\begin{subequations}\label{Eqn:ChannelLaw}
\begin{equation}\label{Eqn:Uplink}
U := X_1 \oplus X_2 \oplus \ldots \oplus X_L \oplus Z,
\end{equation}
where $X_\ell \in \mathcal{F}$ is the symbol sent by user~$\ell$, $U$ is the symbol observed by the relay, and $Z \in \mathcal{F}$ is independent arbitrarily-distributed additive noise. Similarly, the memoryless downlink is 
\begin{equation}\label{Eqn:Downlink}
Y_\ell := V \oplus N_\ell, \quad \ell \in [1,L],
\end{equation}
\end{subequations}
where $V \in \mathcal{F}$ is sent by the relay, $Y_\ell \in \mathcal{F}$ is observed by user~$\ell$, and $N_\ell \in \mathcal{F}$ is independent additive noise at user $\ell$'s receiver. Figure~\ref{Fig:MWRC} depicts the setup for $L=4$ users. The uplink and downlink are memoryless; that means $Z$ and all $N_\ell$'s are independent, and they are each iid over all channel uses.
 
\subsection{Codes for full data exchange}\label{Sec:MWRC:Codes}
 
An $(m,n)$-code for exchanging the users' source data is specified by a collection of mappings 
\begin{equation}\label{Eqn:Code}
\big\{f_{1,t},\ldots,f_{L,t},\phi_t\big\}_{t=1}^n
\quad
 \text{and}
 \quad
 \big\{g_{1},\ldots,g_{L}\big\}, 
\end{equation}
where $f_{\ell,t} : \setW^m_\ell \times \mathcal{F}^{t-1} \to \mathcal{F}$, $\ \phi_t : \mathcal{F}^{t-1} \to \mathcal{F}$ and $g_\ell :
\mathcal{W}_{\ell}^m \times \mathcal{F}^n \to \setW_{1}^m \times \setW_{2}^m \times \cdots \times \setW_{L}^m$. We assume that the nodes operate in the full-duplex mode, and the uplink and the downlink are perfectly synchronised. During the $t$-th channel use, each user~$\ell \in [1,L]$ sends
\begin{equation*}
X_{\ell,t} := f_{\ell,t}(\bm{W}_\ell,Y_{\ell,1},\ldots,Y_{\ell,t-1})
\end{equation*}
over the uplink, and the relay sends 
\begin{equation*}
V_t := \phi_t(U_1,U_2,\ldots,U_{t-1})
\end{equation*}
over the downlink. User~$\ell$ observes $Y_{\ell,t}$ as per \eqref{Eqn:Downlink} and the relay observes $U_{t}$ as per \eqref{Eqn:Uplink}.
After $n$ channel uses, user~$\ell$ has observed $n$ symbols $
\bm{Y}_\ell = (Y_{\ell,1},\ldots,Y_{\ell,n})$ from the downlink. It outputs
\begin{equation*}
\big( \h{\bm{W}}_{\ell,1},\h{\bm{W}}_{\ell,2},\ldots,\h{\bm{W}}_{\ell,L}\big) := g_{\ell}(\bm{W}_\ell,\bm{Y}_\ell),
\end{equation*}
where $\h{\bm{W}}_{\ell,i}$ denotes its reconstruction of $\bm{W}_i$. Let
\begin{equation*}
\Pe := \Pr\left[\ \bigcup_{\ell = 1}^L \Big\{ g_{\ell}(\bm{W}_\ell,\bm{Y}_\ell) \neq \big( \bm{W}_1,\bm{W}_2,\ldots,\bm{W}_L\big) \Big\} \right]
\end{equation*}
denote the average probability of the event that one or more users make a decoding error, for a given code~\eqref{Eqn:Code}. 


\subsection{Optimal source-channel rate $\kappa^*$}

A \emph{source-channel rate} of $\kappa$ channel symbols per source symbol is said to be \emph{achievable} if the following holds: For any $\epsilon > 0$ there exists non-negative integers $m$ and~$n$ (chosen sufficiently large depending on $\epsilon$) together with an $(m,n)$-code~\eqref{Eqn:Code} such that
\begin{equation*}
\kappa = \frac{n}{m} 
\quad
\text{and}
\quad
\Pe \leq \epsilon.
\end{equation*}

\begin{definition}
The \emph{optimal source-channel rate} is 
\begin{equation*}
\kappa^* := \inf\{ \kappa \geq 0 : \kappa \text{ is achievable}\}.
\end{equation*}
\end{definition}

We now present a lower and an upper bounds on $\kappa^*$. For any subset $\set \subseteq [1,L]$, let 
$
W_\set := (W_\ell : \ell \in \set)
$
denote the tuple of source variables with indices in $\set$. 
Let
\begin{equation}
C_\ell := \log|\mathcal{F}| - \max\{ H(Z), H(N_\ell)\}. \label{eq:channel-capacity-equation}
\end{equation}

\begin{theorem}\label{Thm:MWRC:Lower}
\begin{equation*}
\kappa^*  \geq \Psi,
\end{equation*}
where
\begin{equation*}
\Psi := \max_{\ell \in [1,L]}\frac{1}{{C_\ell}}
H(W_{\ellc}|W_{\ell}).
\end{equation*}
\end{theorem}

\begin{IEEEproof}
See Appendix~\ref{App:Proof:Thm:MWRC:Converse}.
\end{IEEEproof}

Let $\mathcal{P}$ denote the set of all joint probability mass functions (pmfs) on $\setW_1 \times\cdots \times \setW_L$, so that any source $(W_1,\ldots,$ $W_L)$ is specified by some $p \in \mathcal{P}$. For brevity, we write $(W_1,\ldots,W_L) \sim p$. Let $\R(p)$ denote the set of all non-negative real-valued tuples $\bm{r} = (r_1,\ldots,r_L)$ satisfying
\begin{equation}\label{Eqn:R}
\sum_{i \in \set} r_i \geq H(W_\set | W_\setC),
\quad 
\forall\ \set \subset [1,L].
\end{equation}
   
\begin{theorem}\label{Thm:MWRC:Upper}
\begin{equation}
\kappa^*
\leq
\min_{\bm{r} \in \R(p)}
\Upsilon(\bm{r}), \label{eq:mwrc-upper}
\end{equation}
where the minimum is attained by a tuple $\bm{r}$ on the boundary of $\R(p)$ and
\begin{equation*}
\Upsilon(\bm{r}) := \max_{\ell \in [1,L]}
\frac{1}{C_\ell}\sum_{i \in \ellc} r_i,
\end{equation*}
\end{theorem}

\begin{IEEEproof}
See Appendix~\ref{App:Proof:Thm:MWRC:Achievability}.
\end{IEEEproof}

Theorem~\ref{Thm:MWRC:Upper} is proved using standalone source and channel codes, and, in this context, $\R(p)$ represents the achievable rate region of the underlying distributed source coding problem. The reader may recognise that $\R(p)$ is closely related to the \emph{Slepian-Wolf rate region}~\cite[Sec.~15.4.2]{coverthomas06}. Indeed, the Slepian-Wolf region for $(W_1,\ldots,W_L) \sim p$ is given by 
\begin{equation}\label{Eqn:SW} 
\left\{ \bm{r} \in \R(p) : \sum_{\ell = 1}^L r_\ell \geq H\big(W_{[1,L]}\big) \right\}.
\end{equation}
In other words, $\R(p)$ is the Slepian-Wolf rate region without the total sum-rate constraint.
Intuitively, the additional sum-rate constraint in~\eqref{Eqn:SW} does not play a role in $\R(p)$ and Theorem~\ref{Thm:MWRC:Upper} because user $\ell$ always has its own source data $\bm{W}_\ell$ as side information. The omission of this constraint is an important characteristic of the rate region $\R(p)$ that shapes much of the following discussion. 


\subsection{When Theorem~\ref{Thm:MWRC:Lower} meets Theorem~\ref{Thm:MWRC:Upper}}



Constraints~\eqref{Eqn:R} in the definition of $\R(p)$ dictate that for any $\bm{r} \in \R(p)$, we must have that $\Upsilon(\bm{r}) \geq \Psi$.
When this inequality is an equality, we have the following lemma:

\begin{lemma}\label{Lem:Ps}
The lower bound in Theorem~\ref{Thm:MWRC:Lower} meets the upper bound in Theorem~\ref{Thm:MWRC:Upper} if and only if there exists a rate-tuple $\bm{r} \in \R(p)$ such that $\Upsilon(\bm{r}) = \Psi$.
Under this condition, $\kappa^* = \Psi$,
and source-channel separation (as specifically described in Appendix~\ref{App:Proof:Thm:MWRC:Achievability})  is optimal.
\end{lemma}

\begin{IEEEproof}
Recall that for any $\bm{r} \in \R(p)$,
\begin{equation*} 
\Psi \step{a}{\leq} \kappa^* \step{b}{\leq} \min_{\bm{r}' \in \R(p)} \Upsilon(\bm{r}') \leq \Upsilon(\bm{r}),
\end{equation*}
where inequalities (a) and (b) follow from Theorems~\ref{Thm:MWRC:Lower} and \ref{Thm:MWRC:Upper} respectively.  
It follows immediately that if $\Upsilon(\bm{r}) = \Psi$, then the lower bound in Theorem~\ref{Thm:MWRC:Lower} meets the upper bound in Theorem~\ref{Thm:MWRC:Upper}, and  $\kappa^* = \Psi$.

Conversely, if the lower and upper bounds meet, then the rate tuple $\bm{r} \in \R(p)$ that minimises $\Upsilon(\bm{r})$ attains the required condition $\Upsilon(\bm{r}) = \Psi$.
 \end{IEEEproof}



The main contribution of this paper is to establish nontrivial sufficient conditions for which there exists an $\bm{r} \in \R(p)$ such that $\Upsilon(\bm{r}) = \Psi$. 


\section{$\Ps$ --- A Class of Sources for Lemma~\ref{Lem:Ps}}\label{Sec:Ps}

The existence of an $\bm{r} \in \R(p)$ satisfying $\Upsilon(\bm{r}) = \Psi$ depends on both the joint pmf $p$ of the source and the entropies of the channel noises in~\eqref{Eqn:ChannelLaw}. Such an $\bm{r}$ can always be found for the following class of sources, irrespective of the particular channel noise entropies: 
\ifx\doublecolumn\undefined
\begin{equation}\label{Eqn:Rate-Region-Equality-Constraint}
\Ps := \Bigg\{ p' \in \mathcal{P} : 
\exists\ \bm{r} \in \R(p') 
\text{ satisfying }
\sum_{i \in \ellc} r_i 
= H\big(W_{\ellc} \big|W_{\ell}\big),\quad \forall\ \ell \in [1,L]\Bigg\}.
\end{equation}
\else
\begin{multline}\label{Eqn:Rate-Region-Equality-Constraint}
\Ps := \Bigg\{ p' \in \mathcal{P} : 
\exists\ \bm{r} \in \R(p') 
\text{ satisfying }\\
\sum_{i \in \ellc} r_i 
= H\big(W'_{\ellc} \big|W'_{\ell}\big),\quad \forall\ \ell \in [1,L]\Bigg\}.
\end{multline}
\fi

\begin{proposition}\label{Cor:Thm:Ps}
If $(W_1,\ldots,W_L) \sim p \in \Ps$, then there exists an $\bm{r} \in \R(p)$ such that $\Upsilon(\bm{r}) = \Psi$. 
\end{proposition}



The class $\Ps$ is a useful regularity condition for Lemma~\ref{Lem:Ps}. 
Here are two simple examples. 

\begin{example}\label{Exa:2Users}
If $L = 2$, then $\Ps = \mathcal{P}$.
\end{example}

\begin{example}
If $(W_1,\ldots,W_L) \sim p$ are independent random variables, then 
$p \in \Ps$.
\end{example}


We now show how one can establish whether an arbitrary source $(W_1,\ldots,W_L) \sim p$ belongs to $\Ps$ by checking whether a specific rate tuple satisfies the conditions in~\eqref{Eqn:Rate-Region-Equality-Constraint}. 
To this end, we re-write the $L$ equalities in~\eqref{Eqn:Rate-Region-Equality-Constraint}  as a linear system:
\begin{equation}\label{Eqn:Rate-Region-Equality-Constraint2}
\bm{r}\ \bm{T} = \bm{h}(p).
\end{equation}
Here we are to solve for the rate vector $\bm{r} = [r_1\ r_2\ \cdots\ r_L]$, where $\bm{T}$ is the fixed $(L \times L)$-matrix
\begin{equation*}
\bm{T} := 
\begin{bmatrix}
0 & 1 & \dotsc & 1\\
1 & 0 &  & 1\\
\vdots & & \ddots & \\
1 & 1 & \dotsc & 0
\end{bmatrix},
\end{equation*}
and 
\begin{equation}\label{Eqn:h}
\bm{h}(p) := [H(W_{1^\text{c}}|W_{1})\ H(W_{2^\text{c}}|W_{2})\ \cdots\ H(W_{L^\text{c}}|W_{L})].
\end{equation}
The matrix $\bm{T}$ has full rank, and we denote the unique solution of~\eqref{Eqn:Rate-Region-Equality-Constraint2} by
\begin{equation}\label{Eqn:Rstar}
\bm{r}^*(p) := \bm{h}(p) \bm{T}^{-1},
\end{equation}
where $\bm{r}^*(p) = [r^*_1\ r^*_2\ \cdots\ r^*_L]$ and
\begin{equation} \label{Eqn:Rstar-individual}
r^*_\ell = 
\frac{\|\bm{h}(p)\|}{L-1} - H(W_{\ellc}|W_{\ell}),
\quad 
\forall\ \ell \in [1,L].
\end{equation}
The conclusion from the above discussion is that we can test whether or not $p \in \Ps$ by numerically checking whether $\bm{r}^*(p) \in \R(p)$. The next lemma follows immediately.

\begin{lemma}\label{Lem:Ps-2}
\begin{equation*}
\Ps = \big\{ p' \in \mathcal{P} : \bm{r}^*(p') \in \R(p') \big\}.
\end{equation*}
\end{lemma}

We now use Lemma~\ref{Lem:Ps-2} to give an example of a source in $\Ps$, and a source that is not in $\Ps$. 
We will see in the next section that all ``balanced'' sources are in $\Ps$.

\begin{example}\label{Exa:BalancedThreeSource}
Let $B_1,B_2,B_3,B_{1,2},B_{1,3}$ and $B_{2,3}$ be independent and uniformly distributed Bernoulli random variables. Suppose that random variable associated with user one, $W_1$, is string of  three bits, $B_1,B_{1,2},$ and $B_{1,3}$, i.e.,
\begin{equation*}
W_1 := (B_1,B_{1,2},B_{1,3}).
\end{equation*}
Similarly, let $W_2  := (B_2,B_{1,2},B_{2,3})$ and $W_3 := (B_3,B_{1,3},B_{2,3})$. If $p$ is the joint pmf of $(W_1,W_2,W_3)$, then 
\begin{equation*}
\R(p) = \left\{(r_1,r_2,r_3) \in \b{R}^3 : 
\begin{array}{ll}
r_\ell \geq 1, & \forall\ \ell \\
r_\ell + r_{\ell'} \geq 3, & \forall\ \ell \neq \ell'
\end{array}
 \right\},
\end{equation*} 
$\bm{r}^*(p) = (3/2,3/2,3/2)$, and therefore $p \in \Ps$.
\end{example}

\begin{example}\label{Exa:UnbalancedThreeSource}
Remove the Bernoulli variables $B_{1,2}$ and $B_{1,3}$ in Example~\ref{Exa:BalancedThreeSource} to obtain $(W_1,W_2,W_3) \sim p$ given by $W_1 := B_1$, $W_2 := (B_2,B_{2,3})$ and $W_3 := (B_3,B_{2,3})$. We have
\begin{equation*}
\R(p) = \left\{(r_1,r_2,r_3) \in \b{R}^3 : 
\begin{array}{l}
r_\ell \geq 1,\quad \forall\ \ell \\
r_1 + r_2 \geq 2\\
r_1 + r_3 \geq 2\\
r_2 + r_3 \geq 3
\end{array}
 \right\},
\end{equation*} 
$\bm{r}^*(p) = (1/2,3/2,3/2)$, and therefore $p \notin \Ps$.
\end{example}

\begin{remark}
If $p \in \Ps$, then Proposition~\ref{Cor:Thm:Ps} guarantees that the lower bound in Theorem~\ref{Thm:MWRC:Lower} meets the upper bound in Theorem~\ref{Thm:MWRC:Upper}. Otherwise (i.e., if $p \notin \Ps$), there is no such guarantee. The upper and the lower bounds can still be tight, depending on the particular source model and channel noises. The following example describes such a situation.
\end{remark}

\begin{example}
Let $T_1$, $T_2$, $T_3$, $T_{1,2}$, $T_{2,3}$, and $T_{1,3}$ be independent random variables with entropies $H(T_1) = H(T_2) = H(T_3) = 1$, $H(T_{1,2}) = H(T_{1,3}) = 3$, and $H(T_{2,3}) = 8$.
Let $L=3$,  $W_1 = (T_1, T_{1,2}, T_{1,3})$, $W_2 = (T_2, T_{1,2}, T_{2,3})$, $W_3 = (T_3, T_{1,3}, T_{2,3})$, and $p$ be a source pmf that satisfies these conditions. This gives $H(W_2,W_3|W_1) = 10$, $H(W_1,W_2|W_3)= H(W_1,W_3|W_2) = 5$, $r_1^* = 0$, $r_2^* = r_3^* = 5$. Clearly, $\bm{r}^*(p) \notin \R(p)$, and thus $p \notin \Ps$. Consider the following two sets of channel parameters:
\begin{enumerate}
\item $C_\ell = 1$ for all $\ell \in [1,3]$: This gives $\Psi = 10$. Choosing $\bm{r} =(1,5,5) \in \R(p)$, we obtain $\Upsilon(\bm{r}) = 10$. The bounds in Theorems~\ref{Thm:MWRC:Lower} and~\ref{Thm:MWRC:Upper} meet for these channel parameters. 
\item $C_1 =10$ and $C_2=C_3=4$: This gives $\Psi=1$. Conditions for $\R(p)$ dictate that $r_2 + r_3 \geq 10$ and $r_1 \geq 1$. This implies $\max \{ r_1 + r_2, r_1 + r_3\} = r_1 + \max \{ r_2, r_3\} \geq r_1 + \frac{r_2+r_3}{2} \geq 6$, and consequently, $\Upsilon(\bm{r}) \geq \max \{\frac{r_2+r_3}{C_1}, \frac{r_1+r_3}{C_2}, \frac{r_1 + r_2}{C_3}\} \geq 1.5 > \Psi$. The bounds do not meet for these channel parameters.
\end{enumerate}
\end{example}


\section{Balanced Sources and the $I$-Measure}\label{Sec:Balanced}


\ifx\doublecolumn\undefined
\else
\begin{figure*}[t]
\centering
\includegraphics[width=0.9\textwidth]{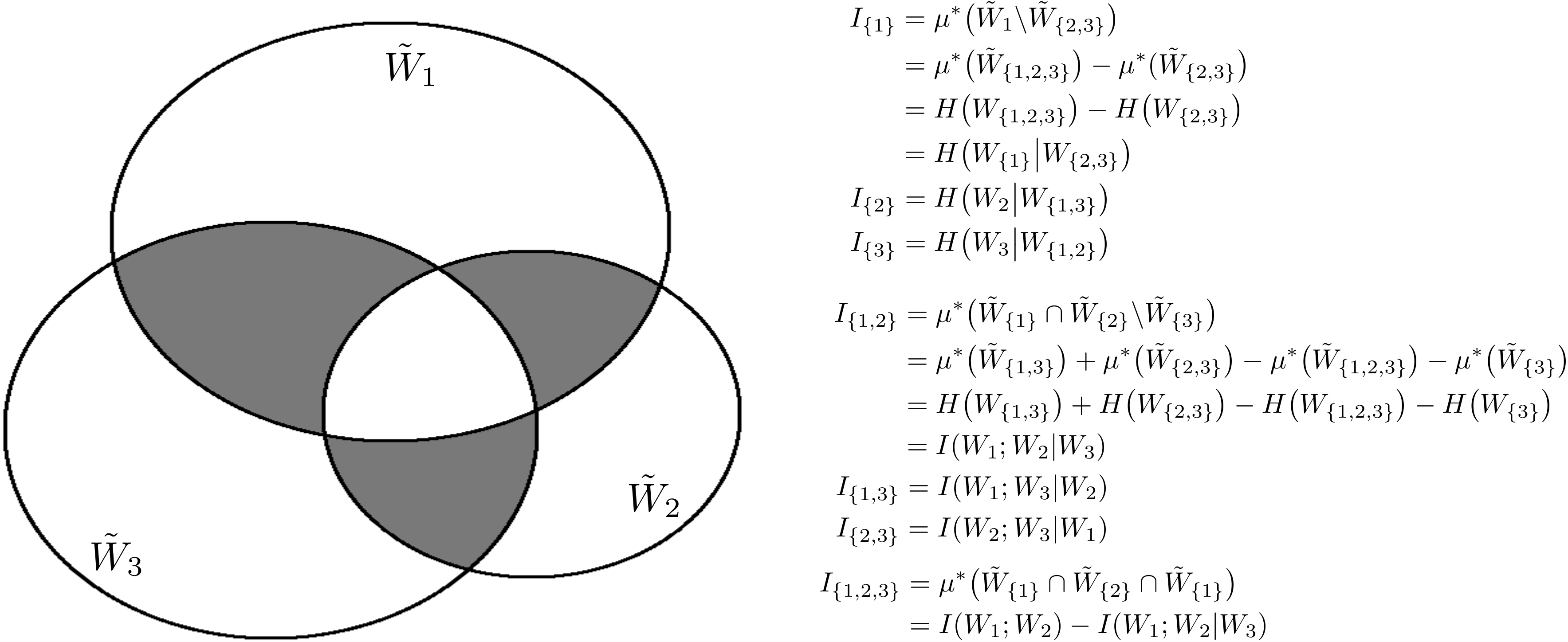}
\caption{Conditional multiple-mutual informations $I_\setK$, $I$-measures $\mu^*$, and information diagram for $(W_1,W_2,W_3)$. The source is balanced if the largest $I$-measure of the shaded areas is not larger that two times the smallest $I$-measure of the shaded areas.}
\label{fig:information-diagram}
\end{figure*}
\fi

\subsection{Balanced sources}

It is sometimes infeasible or undesirable to apply Lemma~\ref{Lem:Ps} by numerically testing whether $\bm{r}^*(p) \in \R(p)$. For example, suppose that we need to verify that a source-channel rate $\kappa$ is achievable for every source within some uncountable set (such a situation is described later in Example~\ref{Exa:CapacityCommonMessages}). In such cases, it is helpful to study more general structural properties of $\Ps$. The next proposition suggests that $\Ps$ is a rather complicated set, and its proof is omitted. 

\begin{proposition}\label{Prop:PsConvex}
$\Ps$ is closed for all $L$, but it is not convex for any $L \geq 3$. 
\end{proposition}

The next proposition was proved by Ong et al.~\cite{onglechnerjohnsonkellett13}. The proposition determines the optimal source-channel rate $\kappa^*$ for a special case of three users and ``balanced mutual information'' sources. 

\begin{proposition}[Ong et al.~{\cite[Thm.~1]{onglechnerjohnsonkellett13}}] \label{Prop:Ong3Users}
If we have $L= 3$ users and the discrete memoryless source $(W_1,W_2,W_3)$ satisfies
\begin{equation}
I(W_i;W_j|W_k) \leq I(W_j;W_k|W_i) + I(W_i;W_k|W_j) \label{eq:L=3}
\end{equation}
for all permutations of $i,j,k \in [1,3]$, then the optimal source-channel rate is given by $\kappa^* = \Psi$.
\end{proposition}

It is relatively easy to prove\footnote{Ong et al.~\cite{onglechnerjohnsonkellett13} gave a direct proof of Proposition~\ref{Prop:Ong3Users} using slightly different techniques.} Proposition~\ref{Prop:Ong3Users} using the ideas in Section~\ref{Sec:Ps}, as shown below:
\begin{IEEEproof}
From~\eqref{Eqn:Rstar-individual}, we get
\ifx\doublecolumn\undefined
\begin{align*}
r^*_1 &:= \frac{1}{2} \Big( H(W_1,W_2|W_3) + H(W_1,W_3|W_2) - H(W_2,W_3|W_1) \Big),\\
r^*_2 &:= \frac{1}{2} \Big( H(W_1,W_2|W_3) + H(W_2,W_3|W_1) - H(W_1,W_3|W_2) \Big),\\
r^*_3 &:= \frac{1}{2} \Big( H(W_1,W_3|W_2) + H(W_2,W_3|W_1) - H(W_1,W_2|W_3) \Big).
\end{align*}
\else
\begin{align*}
r^*_1 &:= \frac{1}{2} \Big( H(W_1,W_2|W_3) + H(W_1,W_3|W_2)\\
&\quad\quad\quad - H(W_2,W_3|W_1) \Big),\\
r^*_2 &:= \frac{1}{2} \Big( H(W_1,W_2|W_3) + H(W_2,W_3|W_1)\\
&\quad\quad\quad  - H(W_1,W_3|W_2) \Big),\\
r^*_3 &:= \frac{1}{2} \Big( H(W_1,W_3|W_2) + H(W_2,W_3|W_1)\\
&\quad\quad\quad  - H(W_1,W_2|W_3) \Big).
\end{align*}
\fi

By construction, we clearly have 
\begin{align*}
r^*_1 + r^*_2 &= H(W_1,W_2|W_3),\\
r^*_1 + r^*_3 &= H(W_1,W_3|W_2),\\
r^*_2 + r^*_3 &= H(W_2,W_3|W_1).
\end{align*}
Moreover, it follows from~\eqref{eq:L=3} that 
\begin{align*}
r^*_1 &\geq H(W_1|W_2,W_3),\\
r^*_2 &\geq H(W_2|W_1,W_3),\\
r^*_3 &\geq H(W_3|W_1,W_2),
\end{align*}
and, therefore, $(r^*_1,r^*_2,r^*_3) \in \R(p)$. This implies that $p \in \Ps$, and Proposition~\ref{Cor:Thm:Ps} gives the desired result.
\end{IEEEproof}

Given the above proof, it is natural to wonder whether one can find a similar ``balanced mutual information'' condition that works more generally for $L \geq 3$. It turns out that such a generalisation is possible, and we now formalise this idea. 



Fix $L \geq 3$ and $(W_1,\ldots,W_L) \sim p$. 
Consider any nonempty subset 
\begin{equation}\label{Eqn:SetK}
\setK = \{\ell_1,\ldots,\ell_k\} \subseteq [1,L].
\end{equation}
The \emph{conditional multiple-mutual information}\footnote{Conditional multiple-mutual information is also called \emph{conditional $k$-information}~\cite{mcgill54,han80b,hekstrawillems89}.} between the random variables $(W_{\ell_1},W_{\ell_2},$ $\ldots,W_{\ell_k})$ was defined by Hekstra and Willems~\cite[Sec.~II.D]{hekstrawillems89}
\ifx\doublecolumn\undefined
\begin{equation*}
I(W_{\ell_1};W_{\ell_2};\cdots ; W_{\ell_k}|W_{\setK^\text{c}})
:=
\sum\limits_{t = 1}^{k} (-1)^{t-1} 
\sum\limits_{\substack{\mathcal{T} \subseteq \setK
\\ \text{ s.t. } |\mathcal{T}| = t}} 
H(W_{\mathcal{T}}|W_{\setK^\text{c}}).
\end{equation*}
\else
\begin{multline*}
I(W_{\ell_1};W_{\ell_2};\cdots ; W_{\ell_k}|W_{\setK^\text{c}})
\\  :=
\sum\limits_{t = 1}^{k} (-1)^{t-1} 
\hspace{-2mm}
\sum\limits_{\substack{\mathcal{T} \subseteq \setK
\\ \text{ s.t. } |\mathcal{T}| = t}} 
H(W_{\mathcal{T}}|W_{\setK^\text{c}}).
\end{multline*}
\fi

In this paper, it will be convenient to define $I_{\emptyset} := 0$ and the notation
\begin{equation*}
I_{\setK} := I(W_{\ell_1};W_{\ell_2};\cdots ; W_{\ell_k}|W_{\setK^\text{c}})
\end{equation*}
for any nonempty subset~\eqref{Eqn:SetK}.

\begin{definition}\label{Def:Balanced}
We say that a source $(W_1,\ldots,W_L) \sim p$ is \emph{balanced}\footnote{The definition of a balanced source here is different from that by Haitner et al.~\cite[Sec.~3]{haitnerhorvitz05}. Here, we consider a source consisting of multiple ``components'', and require that the components $\{W_\ell\}$ have ``roughly'' the same conditional multiple-mutual informations. Haitner et al.'s balance condition is defined for any pmf, and requires that the pmf be ``close to uniform most of the time.''} if 
\begin{equation}\label{eq:balance-condition}
\overline{\mu}_k \leq \mathsf{gap}_k\ \underline{\mu}_k,
\end{equation}
holds for all $k \in [2,L-1]$, where 
\begin{equation*}
\overline{\mu}_k := \max_{\substack{\set \subseteq [1,L] \\ \text{ s.t. } |\set| = k}} I_\set,
\end{equation*}
\begin{equation*}
\underline{\mu}_k := \min_{\substack{\set \subseteq [1,L] \\ \text{ s.t. } |\set| = k}}  I_\set,
\end{equation*}
and
\begin{equation*}
\mathsf{gap}_{k} := 1 + \frac{1}{k} \left( \frac{L-1}{2L-k-3} \right).
\end{equation*}
Let $\mathcal{P}_\text{bal}$ denote the set of all balanced sources.
\end{definition}

\begin{theorem}\label{Thm:Balanced} $\mathcal{P}_\text{bal} \subseteq \Ps$.
\end{theorem}
\begin{IEEEproof}
See Appendix~\ref{App:Proof:Thm:Balanced}.
\end{IEEEproof}

It immediately follows from Theorem~\ref{Thm:Balanced} that the optimal source-channel rate of any balanced source (regardless of the channel noise entropies) is $\kappa^* = \Psi$. While the set $\mathcal{P}_\text{bal}$ may not be as large as $\Ps$, we will see in Section~\ref{Sec:Imeasure} 
that $\mathcal{P}_\text{bal}$ has a measure-theoretic interpretation via the $I$-measure~\cite[Chap.~3]{yeung08}. Consequently, checking condition~\eqref{eq:balance-condition} to determine if a source is balanced is equivalent to comparing different areas in \textit{information diagrams}.


The key idea underlying Definition~\ref{Def:Balanced} is that a balanced source will have 
\begin{equation*}
I_\setK \approx I_{\setK'},
\quad \forall\ \setK,\ \setK' 
\quad \text{with}
\quad |\setK| = |\setK'|.
\end{equation*}
Here the approximation becomes more stringent as the number of users $L$ and the subset cardinalities grow large, in which case the multiplicative factor $\mathsf{gap}_{k}$ in~\eqref{eq:balance-condition} approaches unity from above.\footnote{As a result, we expect the class of $\mathcal{P}_\text{bal}$ to be relatively smaller as $L$ increases.}

Condition~\eqref{eq:balance-condition} for balanced sources suggests certain ``symmetry'' of the source pmf. In particular, if the source pmf $p$ is \textit{symmetrical} in the sense that 
\begin{equation*}
p(w_1,w_2,\dotsc, w_L) = p(w_{\ell_1}, w_{\ell_2}, \dotsc, w_{\ell_L})
\end{equation*}
for all permutations of $\ell_1, \ell_2, \dotsc, \ell_L \in [1,L]$, then $I_\mathcal{K} = I_{\mathcal{K}'}$ for all $\mathcal{K}$ and $\mathcal{K}'$ with $|\mathcal{K}| = |\mathcal{K}'|$. So, a source with a symmetrical pmf is balanced. We extend this idea in the following example:

\begin{example}
Consider a random event $B \in \{0,1\}$ with $\Pr\{B=0\}=\rho$, and three sensors each taking a noisy measurement of the event, $W_\ell = B \oplus E_\ell$ for $\ell \in [1,3]$. Here, $E_\ell \in \{0,1\}$ is the measurement error with $\Pr\{E_\ell=0\} = \sigma_\ell$. If the pmf is symmetrical, i.e., $\sigma_1=\sigma_2=\sigma_3$, then the source $(W_1,W_2,W_3)$ is balanced. In addition, since the balance condition~\eqref{eq:balance-condition} does not require all $\{I_\mathcal{K}: |\mathcal{K}|=k\}$ to be equal, the source is still balanced if $\sigma_\ell$'s are close, e.g., (a) $\rho=0.2, \sigma_1=0.10, \sigma_2=0,12, \sigma_3=0.14$; and (b) $\rho=0.2, \sigma_1=0.40, \sigma_2=0,41, \sigma_3=0.42$. Otherwise, the source is not balanced, e.g., if $\rho=0.2, \sigma_1=0.1, \sigma_2=0,12, \sigma_3=0.2$.
\end{example}

Balanced conditional mutual-informations lead to balanced conditional entropies in~\eqref{Eqn:R} by invoking the next lemma. This lemma plays a key role in the proof of Theorem~\ref{Thm:Balanced}. 

\begin{lemma}\label{Lem:Composition}
\begin{equation}\label{eq:k-information-sum}
H(W_\set|W_\setC) = \sum_{\setK \subseteq \set} I_\setK,
\quad \forall\ \set \subseteq [1,L].
\end{equation}
\end{lemma}
\begin{IEEEproof}
See Appendix~\ref{Sec:Proof:Lem:Composition}.
\end{IEEEproof}


\subsection{Visualising balanced sources with the $I$-measure and information diagrams}\label{Sec:Imeasure}

Definition~\ref{Def:Balanced} and Theorem~\ref{Thm:Balanced} can be visualised using information diagrams and the $I$-measure~\cite[Chap.~3]{yeung08}. Fix $L \geq 3$ and the source $(W_1,\ldots,W_L) \sim p$. In the notation and terminology of Yeung~\cite[Chap.~3]{yeung08}, let us associate an arbitrary set $\tilde{W}_\ell$ to each random variable $W_\ell$. The $I$-measure $\mu^*$ (defined shortly) is a signed measure on these sets that is chosen in a specific way so that all of Shannon's information measures for $(W_1,W_2,\ldots,W_L)$ can be recovered from set-theoretic operations on $\tilde{W}_1,\tilde{W}_2,\ldots,\tilde{W}_L$. More specifically, let $\mathcal{F}_n$ denote the \emph{field}\footnote{The collection of all sets that can be generated from $\tilde{W}_1,\tilde{W}_2,\ldots,\tilde{W}_L$ by applying any sequence of the usual set-theoretic operations, i.e., union, intersection, complement, and difference.} generated by $\tilde{W}_1,\ldots,\tilde{W}_L$. For any $\set \subseteq [1,L]$, let 
\begin{equation*}
\tilde{W}_\set := \bigcup_{\ell \in \set} \tilde{W}_\ell
\end{equation*}
denote the union of all sets with indices in $\set$. The \emph{$I$-measure} $\mu^*$ on $\mathcal{F}_n$ is defined by
\begin{equation*}
\mu^*(\tilde{W}_{\set}) := H(W_\set),\quad  \text{for all non-empty $\set \subseteq [1,L]$}.
\end{equation*}
It turns out that this signed measure is the only measure that agrees with all Shannon's information measures~\cite[Thm.~3.9]{yeung08}. For example, the $I$-measure relates to the mutual information $I(W_1;W_2)$ by
\begin{align*}
I(W_1;W_2)
&= H(W_1) + H(W_2) - H(W_1,W_2) \\
&= \mu^*(\tilde{W}_1) +  \mu^*(\tilde{W}_2) -  \mu^*(\tilde{W}_1 \cup \tilde{W}_2)\\
&= \mu^*(\tilde{W}_1 \cap \tilde{W}_2). 
\end{align*}
Or, more generally, the $I$-measure $\mu^*$ relates to conditional mutual-information via
\begin{equation*}
I_\setK = \mu^*\left( \big(\cap_{\ell \in \setK} \tilde{W}_\ell\big) \backslash \tilde{W}_{\setK^\text{c}} \right).
\end{equation*}
The next example uses $\mu^*$ and information diagrams to visualise balanced sources.

\ifx\doublecolumn\undefined
\begin{figure}[t]
\centering
\includegraphics[width=0.95\textwidth]{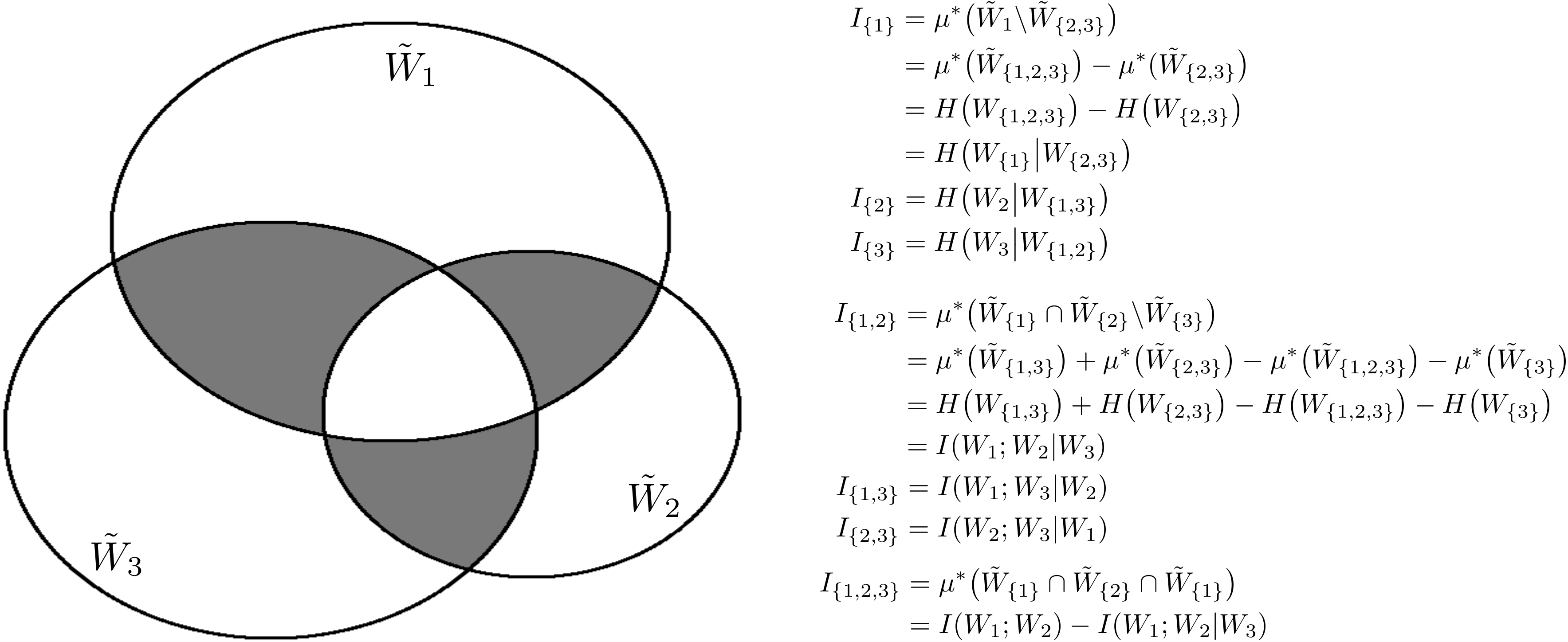}
\caption{Conditional multiple-mutual informations $I_\setK$, $I$-measures $\mu^*$, and information diagram for $(W_1,W_2,W_3)$. The source is balanced if the largest $I$-measure of the shaded areas is not larger that two times the smallest $I$-measure of the shaded areas.}
\vspace{-5mm}
\label{fig:information-diagram}
\end{figure}
\else
\fi

\begin{example}
Consider $L = 3$ users and an arbitrary source $(W_1,W_2,W_3)$. Figure~\ref{fig:information-diagram} depicts the corresponding information diagram, and it lists the values of $\mu^*$ and conditional mutual-information for all subsets of $\{1,2,3\}$. Definition~\ref{Def:Balanced} concerns the $I$-measures of 
\begin{equation*}
(\tilde{W}_1 \cap \tilde{W}_2) \backslash \tilde{W}_3,
\quad (\tilde{W}_1 \cap \tilde{W}_3) \backslash \tilde{W}_2,
\quad \text{and}
\quad  (\tilde{W}_2 \cap \tilde{W}_3) \backslash \tilde{W}_1,
\end{equation*}
which are shaded in Figure~\ref{fig:information-diagram}.
In particular, we have
\begin{align*}
\overline{\mu}_2 
&= \max\big\{ I(W_1;W_2|W_3),I(W_1;W_3|W_2),I(W_2;W_3|W_1)\big\},\\
\underline{\mu}_2 
&=\min\big\{ I(W_1;W_2|W_3),I(W_1;W_3|W_2),I(W_2;W_3|W_1)\big\},
\end{align*}
and the source is balanced if $\overline{\mu}_2 \leq 2 \underline{\mu}_2$.
\end{example}

\subsection{Source-channel rate $\kappa$ and an achievable rate region}

The next example shows how one can use Lemma~\ref{Lem:Ps} and Theorem~\ref{Thm:Balanced} to obtain \emph{achievable rates} for the multiway relay channel with common messages. 

\begin{example}\label{Exa:CapacityCommonMessages}
Suppose that we have $L = 3$ users. For each nonempty $\set \subset [1,3]$, let 
\begin{equation*}
B_\set \in \left[1, 2^{R_\set} \right]
\end{equation*} 
be an independent and uniformly-distributed random variable, where $R_\set \geq 0$ and $2^{R_\set}$ is an integer. Fix the source-channel rate $\kappa>0$, and let $(W_1,W_2,W_3) \sim p$ be given by 
\begin{align*}
W_1 &:= (B_{\{1\}},B_{\{1,2\}},B_{\{1,3\}}),\\
W_2 &:= (B_{\{2\}},B_{\{1,2\}},B_{\{2,3\}}),\\
W_3 &:= (B_{\{3\}},B_{\{1,3\}},B_{\{2,3\}}).
\end{align*}
We wish to characterise the set of all tuples $(R_\set : \set \subset [1,3])$ for which $\kappa$ is achievable.

We have $I_{\set} = R_{\set}$ for all $\set \subset [1,3]$. We say that the rate tuple $(R_\set : \set \subset [1,3])$ is balanced if the corresponding source is balanced, that is, when
\begin{equation}
\frac{\max\big\{R_{\{1,2\}},R_{\{1,3\}},R_{\{2,3\}}\big\}}{\min\big\{R_{\{1,2\}},R_{\{1,3\}},R_{\{2,3\}}\big\}}
 \leq 2. \label{eq:example-balanced}
\end{equation}

Applying Lemma~\ref{Lem:Ps} and Theorem~\ref{Thm:Balanced}, we have that $\kappa$ is achievable for a balanced rate tuple $(R_\set : \set \subset [1,3])$  if (and only if)\footnote{Replace the strict inequality $>$ with an inequality for the case of only if.}
\begin{align*}
\kappa 
> 
\max &\left\{
\frac{R_{\{2\}} + R_{\{3\}} + R_{\{2,3\}} }{\log|\mathcal{F}| - \max\{H(Z),H(N_1)\}},
\right.\\
%
&\hspace{2.2mm}\left. \frac{R_{\{1\}} + R_{\{3\}} + R_{\{1,3\}} }{\log|\mathcal{F}| - \max\{H(Z),H(N_2)\}} \right.,\\
%
&\hspace{2.2mm}\left.
\frac{R_{\{1\}} + R_{\{2\}} + R_{\{1,2\}} }{\log|\mathcal{F}| - \max\{H(Z),H(N_3)\}}\right\}.
\end{align*} 

\end{example}

   

\section{Beyond Relaying: An Application of $\Ps$ to Centralised Storage Systems}\label{Sec:Storage} 
 
 The ideas in Sections~\ref{Sec:Ps} and~\ref{Sec:Balanced} concern the class of sources $\Ps$, and they can be applied to any multiterminal problem for which $\R(p)$ is a meaningful rate region. To illustrate this idea, we now present one such example concerning the centralised storage of correlated data. 


\subsection{Problem setup}

\begin{figure}[t]
\centering
\includegraphics[width=0.35\textwidth]{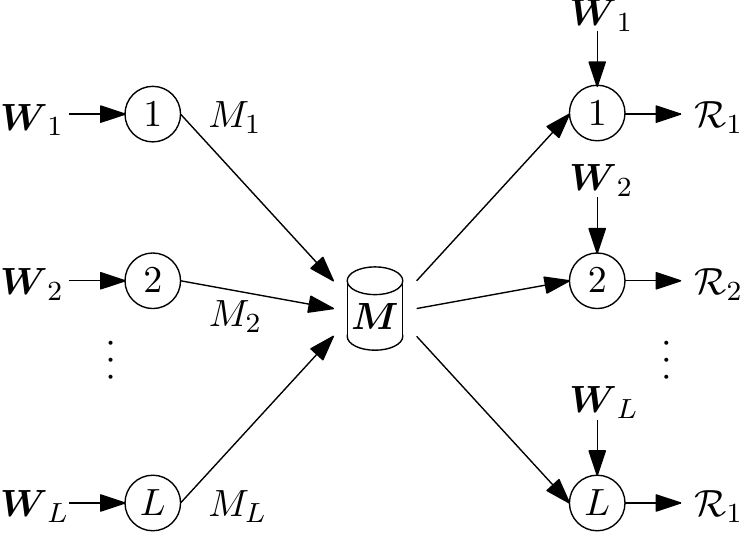}
\caption{A centralised storage system with $L$ clients.}
\label{fig:storage}
\end{figure}

Consider the data-storage system depicted in Figure~\ref{fig:storage}. The $L$ clients
have correlated source data that they wish to write to the storage device. Suppose that the clients' data is generated by the source in Section~\ref{Sec:MWRC}, and that the data of client $\ell$ is the iid $m$-tuple $\bm{W}_\ell = (W_{\ell,1},W_{\ell,2},\ldots,W_{\ell,m}).$ We assume that the method of storage must allow \emph{any} client in the future to reliably recover the source data of \emph{any} subset of clients. We also assume that the storage device is ``dumb'' in the sense that the clients can read and write, but the device itself does not process the stored data. 

Client $\ell$ writes $M_\ell = f_i(\bm{W}_\ell)$ to the storage device, where $f_\ell : \setW_\ell^m \to \setM_\ell$. At some future time, client $\ell$ will attempt to recover the source data of all other users source data by computing 
\begin{equation*}
\big(\hat{\bm{W}}_{\ell,1},\hat{\bm{W}}_{\ell,2},\ldots,\hat{\bm{W}}_{\ell,L} \big) := g_\ell(\bm{W}_\ell,M_1,M_2,\ldots,M_L),
\end{equation*}
where $g_\ell : \setW_\ell \times \setM_1 \times \cdots \times \setM_L \to \setW_1^m \times \cdots \times \setW_L^m$. We call $(f_1,\ldots,f_L,g_1,\ldots,g_L)$ an $m$-code. Let
\begin{equation*}
\Pe := \Pr\left[\ \bigcup_{\ell = 1}^L \Big\{ g_{\ell}(M_1,\ldots,M_L) \neq \big( \bm{W}_1,\ldots,\bm{W}_L\big) \Big\} \right]
\end{equation*}
denote the code's average probability of error, and let
\begin{equation*}
r_{\Sigma} := \sum_{\ell = 1}^L \frac{1}{m} \log |\setM_\ell|
\end{equation*}
denote the total \emph{storage rate} needed by the device (that means the device needs at least $m r_\Sigma$ bits to store the clients' data).


\subsection{Optimal storage rate}

We say that a \emph{storage rate} of $r_\Sigma$ is \emph{achievable} if for any $\epsilon > 0$ there exists a sufficiently large $m$ and $m$-code where $\Pe \leq \epsilon$.
The \emph{optimal storage rate} is 
\begin{equation*}
r_\Sigma^* := \inf\big\{ r_\Sigma \geq 0 : r_\Sigma \text{ is achievable}\big\}.
\end{equation*}

The next theorem is proved in Appendix~\ref{Sec:Proof:Thm:Storage}. 


\begin{theorem}\label{Thm:Storage}
Considering $(W_1,\dots,W_L) \sim p$,
\begin{equation}\label{Eqn:Thm:SourceCoding}
r_\Sigma^* = \min_{\bm{r} \in \R(p)}\ \|\bm{r}\|.
\end{equation}
\end{theorem}

The next corollary specialises Theorem~\ref{Thm:Storage} to give a closed-form expression for the optimal storage rate of any source in $\Ps$.

\begin{corollary}
If $(W_1,\dots,W_L) \sim p$ with $p \in \Ps$, then
\begin{equation*}
r_\Sigma^* = \frac{1}{L-1} \|\bm{h}(p)\|,
\end{equation*} 
where 
\begin{equation*}
\bm{h}(p) = [H(W_{1^\text{c}}|W_{1})\ H(W_{2^\text{c}}|W_{2})\ \cdots\ H(W_{L^\text{c}}|W_{L})].
\end{equation*}
\end{corollary}

\begin{IEEEproof}
Recall the rate tuple $\bm{r}^*(p) = [r^*_1,\ldots,r^*_L]$ in~\eqref{Eqn:Rstar} defined by  
\begin{equation*}
r^*_\ell = \frac{\|\bm{h}(p)\|}{L-1} - H(W_\ellc | W_\ell),
\quad 
\forall\ \ell \in [1,L].
\end{equation*}
Lemma~\ref{Lem:Ps-2} showed that $\bm{r}^*(p) \in \R(p)$ if and only if $p \in \Ps$. We will next show that the rate tuple $\bm{r}^*(p)$ attains the right hand side of~\eqref{Eqn:Thm:SourceCoding}, and therefore we have
\begin{equation}\label{Eqn:Proof:Cor:Thm:Storage}
r_\Sigma^*  = \|\bm{r}^*(p)\|.
\end{equation}
To see why~\eqref{Eqn:Proof:Cor:Thm:Storage} must be true, let us suppose, to the contrary, that there exists another $\bm{r}' \in \R(p)$ with $\|\bm{r}'\| < \|\bm{r}^*(p)\|$. By construction, $\bm{r}^*(p)$ is the unique solution of the linear system
\begin{equation*}
\sum_{i \in \ellc} r^*_i = H(W_\ellc|W_\ell),\quad \forall\ \ell \in [1,L].
\end{equation*}  
Hence, there exists an $\ell \in [1,L]$ such that $\bm{r}'$ satisfies 
\begin{equation*}
\sum_{i \in \ellc} r'_i < \sum_{i \in \ellc} r^*_i = H(W_\ellc|W_\ell).
\end{equation*}
This strict inequality leads to the contradiction $\bm{r}' \notin \R(p)$, and thus $\bm{r}^*(p)$ must achieve the minimum in~\eqref{Eqn:Thm:SourceCoding}. Finally, \eqref{Eqn:Proof:Cor:Thm:Storage} simplifies to
\begin{align*}
r_\Sigma^*
&= \|\bm{r}^*(p)\| \\
&=
\sum_{\ell = 1}^L \left( \frac{\|\bm{h}(p)\|}{L-1} - H(W_\ellc|W_\ell) \right)\\
&=
\frac{L}{L-1} \|\bm{h}(p)\| -  \|\bm{h}(p)\|\\
&= \frac{\|\bm{h}(p)\|}{L-1}. \tag*{\IEEEQEDhere\ }
\end{align*}
\end{IEEEproof}


\section{Summary and Conclusions}
Finding the optimal source-channel rate $\kappa^*$ of the multiway relay channel is an open problem whose solution will provide design insights for cooperative communications systems. We presented simple lower and upper bounds on $\kappa^*$ in Theorems~\ref{Thm:MWRC:Lower} and~\ref{Thm:MWRC:Upper}. In Lemma~\ref{Lem:Ps}, we leveraged these bounds to give a closed-form expression for $\kappa^*$ and a source-channel separation theorem.

Lemma~\ref{Lem:Ps} holds for all combinations of sources and channels where there exists a rate tuple $\bm{r} \in \R(p)$ such that $\Upsilon(\bm{r}) = \Psi$ (that is, the lower bound in Theorem~\ref{Thm:MWRC:Lower} meets the upper bound in Theorem~\ref{Thm:MWRC:Upper}).  
Here $\R(p)$ is the achievable rate region of the underlying distributed source-coding problem, and the condition $\Upsilon(\bm{r}) = \Psi$ depends on both the source and the channel. In general, it remains an open problem to determine $\kappa^*$ for  source-channel combinations where there does not exist such an $\bm{r}$, and for these source-channel combinations, it may be useful to bound the gap between Theorems~\ref{Thm:MWRC:Lower} and~\ref{Thm:MWRC:Upper}.

Unfortunately, it can be difficult to determine when Lemma~\ref{Lem:Ps} holds, and 
for this reason we presented two regularity conditions in Sections~\ref{Sec:Ps} and~\ref{Sec:Balanced}. The first regularity condition describes a class of sources $\Ps$ for which Lemma~\ref{Lem:Ps} is guaranteed to hold, regardless of the channel. Testing whether or not a given source belongs to $\Ps$ involves solving a linear system (see~\eqref{Eqn:Rstar}). The second regularity condition describes a class of balanced sources $\mathcal{P}_\text{bal} \subseteq \Ps$ using conditional multiple-mutual informations. This balance condition can be easily understood via the $I$-measure and information diagrams, and it is most useful in problems where the source is specified by its $I$-measures (see Example~\ref{Exa:CapacityCommonMessages}). 

Finally, the source classes $\Ps$ and $\mathcal{P}_\text{bal}$ concern only the entropic structure of the distributed source coding rate region $\R(p)$ and, therefore, can be applied to any problem where $\R(p)$ is meaningful. To illustrate this idea, we used $\Ps$ and $\mathcal{P}_\text{bal}$ to describe an optimal storage rate for a centralised storage problem  in Section~\ref{Sec:Storage}.  


\appendices 

\section{Proof of Theorem~\ref{Thm:MWRC:Lower}}\label{App:Proof:Thm:MWRC:Converse}

Suppose that $\kappa > 0$ is achievable ($\kappa = 0$ is trivial). Fix $0 < \epsilon \leq 1/2$. There exists  integers $m$ and $n$ with $n/m = \kappa$ and an $(m,n)$-code~\eqref{Eqn:Code} satisfying $\Pe \leq \epsilon$ for any $\epsilon > 0$. For any $\ell \in [1,L]$,
\begin{align}
m H(W_\ellc | W_\ell) 
\notag
&\step{a}{=} 
H(\bm{W}_\ellc | \bm{W}_\ell),\\
\notag
&\step{b}{\leq} 
H(\bm{W}_\ellc | \bm{W}_\ell) 
- H(\bm{W}_\ellc | \bm{W}_\ell,\bm{Y}_\ell) 
+ \varepsilon(m,\epsilon),\\
%
%
%
%
\notag
&\step{c}{\leq} I(\bm{V};\bm{Y}_\ell)
+ \varepsilon(m,\epsilon)\\
\notag
&\step{d}{\leq}
\sum_{i = 1}^n
I(V_i; Y_{\ell,i})
+ \varepsilon(m,\epsilon)\\
\label{Eqn:Thm:main:LB1}
&\step{e}{\leq}
n \big(\log |\mathcal{F}| - H(N_\ell) \big) + \varepsilon(m,\epsilon).
\end{align}
Notes on~\eqref{Eqn:Thm:main:LB1}:
\begin{enumerate}
\item[a.] The source is iid.

\item[b.] Noting that $\Pe \leq \epsilon \leq 1/2$ and invoking Fano's inequality~\cite[Thm.\ 2.10.1]{coverthomas06}, we get  $H(\bm{W}_\ellc | \bm{W}_\ell,\bm{Y}_\ell) \leq \varepsilon(m,\epsilon)$, where
\begin{equation*}
\varepsilon(m,\epsilon) := 1 + \epsilon m \sum_{i \in \ellc} \log |\setW_i|. 
\end{equation*} 

\item[c.] By the chain rule and non-negativity of conditional mutual information, we have
\begin{align*}
I(\bm{W}_\ellc; \bm{Y}_\ell| \bm{W}_\ell)
&\leq 
I(\bm{W}_\ellc,\bm{V}; \bm{Y}_\ell| \bm{W}_\ell)\\
&= 
I(\bm{W}_\ellc,\bm{W}_\ell,\bm{V}; \bm{Y}_\ell)
- I(\bm{W}_\ell; \bm{Y}_\ell)\\
&\leq
I(\bm{V}; \bm{Y}_\ell),
\end{align*}
where the last inequality follows because $(\bm{W}_\ellc,\bm{W}_\ell) \leftrightarrow \bm{V} \leftrightarrow \bm{Y}_\ell$ forms a Markov chain.

\item[d.] The downlink channel is memoryless; in particular,
\begin{align*}
I(\bm{V};\bm{Y}_\ell)
&= \sum_{i=1}^n I(\bm{V};Y_{\ell,i}|Y_{\ell,1},\ldots,Y_{\ell,i-1})\\
&\leq 
\sum_{i=1}^n 
\Big( H(Y_{\ell,i}) - H(Y_{\ell,i}|\bm{V},Y_{\ell,1}^{i-1})\Big)\\
&\step{d.1}{=}
\sum_{i=1}^n \Big(H(Y_{\ell,i}) - H(Y_{\ell,i}|V_i)\Big),
\end{align*}
where (d.1) follows because $Y_{\ell,i} 
\leftrightarrow V_i \leftrightarrow (V_{1}^{i-1},V_{i+1}^n,Y_{\ell,1}^{i-1})$ forms a Markov chain.

\item[e.] $I(V_i;Y_{\ell,i}) = H(Y_{\ell,i}) - H(Y_{\ell,i}|V_i) \leq \log|\mathcal{F}| - H(N_{\ell})$, 
where $ H(Y_{\ell,i}) \leq \log |\mathcal{F}|$, and $H(Y_{\ell,i}|V_i) = H(N_\ell)$ follows from the additive-noise channel law $Y_{\ell,i} = V_i \oplus N_\ell$.
\end{enumerate}

By similar arguments, we have 
\begin{equation}\label{Eqn:Thm:main:LB2}
m H(W_\ellc | W_\ell) 
\leq
n \big(\log|\mathcal{F}| - H(Z) \big)+ \varepsilon(m,\epsilon),
\end{equation}
%
Combining $\kappa = n/m$ together with~\eqref{Eqn:Thm:main:LB1} and~\eqref{Eqn:Thm:main:LB2}, we get
\begin{equation}\label{Eqn:Thm:main:LB3}
\kappa \geq \frac{H(W_\ellc|W_\ell)-\varepsilon(m,\epsilon)/m}{\log|\mathcal{F}| - \max\{ H(Z),H(N_\ell)\}},
\quad 
\forall\ \ell \in [1,L].
\end{equation}
For any $\epsilon>0$, since \eqref{Eqn:Thm:main:LB3} must hold for all sufficiently large $m$,  we must have
\begin{equation*}
\kappa \geq \frac{H(W_\ellc|W_\ell)}{\log|\mathcal{F}| - \max\{ H(Z),H(N_\ell)\}},
\quad 
\forall\ \ell \in [1,L]. \tag*{$\blacksquare$}
\end{equation*}


\section{Proof of Theorem~\ref{Thm:MWRC:Upper}}\label{App:Proof:Thm:MWRC:Achievability}

We use the standard (strict sense) separate source-channel coding technique to prove Theorem~\ref{Thm:MWRC:Upper}. The channel capacity and source-coding regions of interest are defined next. 

\subsection{Channel capacity region}

For each $\ell \in [1,L]$, let $M_\ell \in \setM_\ell$ be an independent and uniformly distributed random variable (a channel-coding message) on a finite set $\setM_\ell$. Recast the joint source-channel coding problem in Section~\ref{Sec:MWRC} as a pure channel coding problem with $M_\ell$ in place of $\bm{W}_\ell$. More specifically, we
\begin{itemize}
\item define an $n$-code via~\eqref{Eqn:Code} by setting $m = 1$ and replacing $\bm{W}_\ell$ with $M_\ell$ and $\setW_\ell$ with $\setM_\ell$ throughout Section~\ref{Sec:MWRC:Codes}, and
\item require each user to exchange its message with that of every other user. 
\end{itemize}
For any given $n$-code, let
\begin{equation*}
\Pe := \Pr\left[ \bigcup_{\ell = 1}^L \Big\{ g_\ell(M_\ell,\bm{Y}_\ell) \neq (M_1,M_2,\ldots,M_L)\Big\} \right]
\end{equation*}
denote the average probability of error, and let $\bm{R} = (R_1, R_2, \ldots, R_L)$ with
\begin{equation*}
R_\ell := \frac{1}{n} \log_2 |\setM_\ell|
\end{equation*}
denote the channel-coding rates of each user (in bits per channel use).

A channel-coding rate tuple $\bm{R}$ is \emph{achievable} if for any $\epsilon > 0$ there exists an $n$-code such that $\Pe \leq \epsilon$.
The \emph{capacity region} $\mathcal{C}$ is the closure of set of all achievable rate tuples. 

\begin{lemma}[Ong et al.~\cite{ongmjohnsonit11}]\label{Lem:CapacityRegion}
\begin{equation*}
\mathcal{C} = 
\Bigg\{ \bm{R} \in [0,\infty)^L : 
\sum_{i \in\ellc} R_i \leq C_\ell,
\ \forall\ \ell \in [1,L]\Bigg\},
\end{equation*}
where $C_\ell$ is defined in \eqref{eq:channel-capacity-equation}.
\end{lemma}


\subsection{Source coding region}\label{Sec:SourceCoding}

Consider an arbitrary source $(W_1,\ldots,W_L) \sim p$ and recall the setup of Section~\ref{Sec:MWRC:Source}. Suppose that the users are required to exchange their source data via rate-limited noiseless channels, instead of the noisy finite-field channel. In particular, suppose that user $\ell$ compresses its source data $\bm{W}_\ell$ to a discrete index $M_\ell := f_\ell(\bm{W}_\ell)$, where $f_\ell : \setW_\ell^m \to \setM_\ell$. User $\ell$ is given every index and it attempts to reconstruct the source data of all users:
\begin{equation*}
\big( \h{\bm{W}}_{\ell,1},\h{\bm{W}}_{\ell,2},\ldots,\h{\bm{W}}_{\ell,L}\big) := g_{\ell}(\bm{W}_\ell,M_1,M_2,\ldots,M_L),
\end{equation*}
where $g_\ell : \setW_\ell^m \times \setM_1 \times \cdots \times \setM_L  \to \setW_1^m \times \cdots \times \setW_L^m$. We call the above collection of compressors and decompressors an $m$-code. For any given $m$-code, let 
\begin{equation*}
\Pe := \Pr\left[ \bigcup_{\ell = 1}^L \Big\{ g_\ell(\bm{W}_\ell,M_1,\ldots,M_L) \neq (\bm{W}_1,\ldots,\bm{W}_L)\Big\} \right]
\end{equation*}
denote the average probability of error, and let $\bm{r} = (r_1,r_2,\ldots,r_L)$ with
\begin{equation*}
r_\ell := \frac{1}{m} \log_2 |\setM_\ell|
\end{equation*}
denote the source-coding rates of each user (in bits per source symbol).

A source-coding rate $\bm{r}$ is \emph{achievable} if for any $\epsilon > 0$ there exists an $m$-code such that $\Pe \leq \epsilon$.
The \emph{source coding region} is the closure of the set of all achievable rate tuples. 

\begin{lemma}[Timo et al.\!~\cite{timolechnerongjohnson12}]
Let $(W_1,\ldots,$ $W_L) \sim p$. The source coding region is equal to $\R(p)$.
\end{lemma}


\subsection{Source-channel coding with standalone codes}

Let us now return to the joint source-channel coding problem. Denote the interiors of  $\mathcal{C}$ and $\R(p)$ respectively by
\begin{equation*}
\text{int}(\mathcal{C}) := \big\{\bm{a} \in \mathcal{C} : \exists\ \epsilon > 0 \text{ with } \bm{a} +\mathcal{B}_\bm{\bm{a}}(\epsilon) \subset \mathcal{C} \big\} 
\end{equation*}
and
\begin{equation*}
\text{int}(\R(p)) := \big\{\bm{b} \in \R(p) : \exists\ \epsilon > 0 \text{ with } \bm{b} +\mathcal{B}_\bm{\bm{b}}(\epsilon) \subset \R(p) \big\}, 
\end{equation*}
where $\mathcal{B}_\bm{a}(\epsilon) := \{ \bm{b} \in \mathbb{R}^L : \|\bm{a} - \bm{b}\| \leq \epsilon\}$.
We now prove the first assertion of Theorem~\ref{Thm:MWRC:Upper}. Map the output of user~$\ell$'s source encoder, i.e., $M_\ell \in [1,2^{mr_\ell}]$, to the input of its channel encoder. This mapping is bijective if and only if $R_\ell = r_\ell/\kappa$.
If
\begin{equation}
\bm{R} = \bm{r}/\kappa \in \text{int}(\mathcal{C})\label{eq:separation-1}
\quad
\text{and}
\quad
\bm{r} \in \text{int}(\R(p)),
\end{equation}
then each user can separately perform source and channel decoding, to reliably decode its required message. This means the source-channel rate $\kappa$ is achievable.

We now show that if
\begin{equation}
\kappa > \min_{\bm{r} \in \R(p)} \max_{\ell \in [1,L]} \frac{1}{C_\ell} \sum\limits_{i \in \ellc} r_i, \label{eq:chosen-kappa}
\end{equation}
then there exists a rate tuple $\bm{r}$ such that \eqref{eq:separation-1} holds, and therefore $\kappa$ is achievable.

Firstly, let $\bm{r}^\dagger=(r^\dagger_1, r^\dagger_2, \dotsc, r^\dagger_L)$ be a rate tuple that attains $\min_{\bm{r} \in \R(p)} \max_{\ell \in [1,L]}\frac{1}{C_\ell} \sum_{i \in \ellc} r_i$. This means the chosen $\kappa$ in \eqref{eq:chosen-kappa} can be written as
\begin{subequations}
\begin{align}
\kappa &= \delta + \max_{\ell \in [1,L]} \frac{1}{C_\ell} \sum\limits_{i \in \ellc} r^\dagger_i\\
& > \max_{\ell \in [1,L]} \frac{1}{C_\ell} \sum\limits_{i \in \ellc} (r^\dagger_i + \rho)\\
& = \max_{\ell \in [1,L]}\frac{1}{C_\ell} \sum\limits_{i \in \ellc} r'_i,
\end{align}
\end{subequations}
for some $\delta >0$, where $\rho := \frac{\delta \min_{\ell \in [1,L]} C_\ell }{L} > 0$, and $r'_i := r^\dagger_i + \rho$.

Now, let $\bm{r}' = (r'_1, r'_2, \dotsc, r'_L)$. Clearly, since $\bm{r}^\dagger \in \R(p)$, we have $\bm{r}' \in \text{int}(\R(p))$. Also, for each $\ell \in [1,L]$, we select
\begin{equation*}
R'_\ell = \frac{r'_\ell}{\kappa} < \frac{r'_\ell}{\max_{k \in [1,L]}\frac{1}{C_k} \sum_{i \in k^\text{c}} r'_i}. 
\end{equation*}
It follows that, for each $\ell \in [1,L]$,
\begin{equation*}
\sum_{j \in \ellc} R'_j <  \frac{\sum_{j \in \ellc} r'_j}{\max_{k \in [1,L]} \frac{1}{C_k} \sum_{i \in k^\text{c}} r_i'}
\leq \frac{\sum_{j \in \ellc} r'_j}{\frac{1}{C_\ell}} \sum_{j \in \ellc} r_j' = C_\ell.
\end{equation*}
This means $\bm{r}'/\kappa \in \text{int}(\mathcal{C})$. Since any $\kappa$ satisfying \eqref{eq:chosen-kappa} is achievable, we have \eqref{eq:mwrc-upper}.

Finally, since the region $\R(p)$ is closed, and $\max_{k \in [1,L]} \frac{1}{C_k} \sum_{i \in k^\text{c}} r_i$ is a strictly-increasing function of any $r_i$, the right-hand side of \eqref{eq:mwrc-upper} is attained by a tuple $\bm{r}$ on the boundary of $\R(p)$. This completes the proof of Theorem~\ref{Thm:MWRC:Upper}.  \hfill $\blacksquare$

\section{Proof of Lemma~\ref{Lem:Composition}}\label{Sec:Proof:Lem:Composition}

We now show that 
\begin{equation}\label{Eqn:Lem:Composition}
H(W_\set|W_{\set^\text{c}}) 
= 
\sum_{\setK \subseteq \set}
I_\setK,\quad \forall\ \set \subseteq [1,L].
\end{equation}
The proof follows by induction: We first show that~\eqref{Eqn:Lem:Composition} holds for all subsets with cardinality $0$ and $1$. We then show that the truth of~\eqref{Eqn:Lem:Composition} for any subset $\set \subset [1,L]$ implies the truth of~\eqref{Eqn:Lem:Composition} for all subsets $\set' \subseteq [1,L]$ of cardinality $|\set'| = |\set| + 1$.  

Starting with cardinality $0$ and the empty set, we have 
\begin{equation}\label{Eqn:Proof:Lem:Composition1}
\sum_{\setK \subseteq \emptyset}
I_\setK 
= 
I_\emptyset 
=
H(W_\emptyset|W_{[1,L]}) = 0. 
\end{equation}
Now consider any singleton $\{\ell\} \subset [1,L]$. We have
\begin{equation}\label{Eqn:Proof:Lem:Composition2}
\sum_{\setK \subseteq \{\ell\}} I_\setK 
= 
I_\emptyset + I_{\{\ell\}}
= H(W_\ell|W_\ellc).  
\end{equation} 

Suppose now that we are given $\set \subset [1,L]$ such that 
\begin{equation}\label{Eqn:Proof:Lem:Composition3}
\sum_{\setK \subseteq \set} I_\setK = H(W_{\set} | W_{\set^\text{c}}).
\end{equation}
For any $j \in \set^\text{c}$, we have
\ifx\doublecolumn\undefined
\begin{align}
H(W_{\set \cup \{j\}} | W_{(\set \cup \{j\})^\text{c}})
\notag
&= 
H(W_{\set \cup \{j\}} | W_{\set^\text{c} \backslash\{j\}})\\
\notag
&\step{a}{=}
H(W_{\set} | W_{\set^\text{c} \backslash\{j\}},W_j) 
+ H(W_j | W_{\set^\text{c} \backslash\{j\}})\\
\notag
&\step{b}{=}
\sum_{\setK \subseteq \set} I_\setK
+ H(W_j | W_{\set^\text{c} \backslash\{j\}})\\
\notag
&\step{c}{=}
\sum_{\setK \subseteq \set} I_\setK
+\sum_{\setK \subseteq \set} I_{\setK \cup \{j\}} \\
\label{Eqn:Proof:Lem:Composition4}
&=
\sum_{\setK \subseteq \set \cup \{j\}} I_\setK,
\end{align} 
\else
\begin{align}
H(W_{\set \cup \{j\}} | W_{(\set \cup \{j\})^\text{c}})
\notag
&= 
H(W_{\set \cup \{j\}} | W_{\set^\text{c} \backslash\{j\}})\\
\notag
&\step{a}{=}
H(W_{\set} | W_{\set^\text{c} \backslash\{j\}},W_j) \\
\notag
&\quad
+ H(W_j | W_{\set^\text{c} \backslash\{j\}})\\
\notag
&\step{b}{=}
\sum_{\setK \subseteq \set} I_\setK
+ H(W_j | W_{\set^\text{c} \backslash\{j\}})\\
\notag
&\step{c}{=}
\sum_{\setK \subseteq \set} I_\setK
+\sum_{\setK \subseteq \set} I_{\setK \cup \{j\}} \\
\label{Eqn:Proof:Lem:Composition4}
&=
\sum_{\setK \subseteq \set \cup \{j\}} I_\setK,
\end{align} 
\fi
where step (a) applies the chain rule for entropy, and step (b) follows by the inductive assumption~\eqref{Eqn:Proof:Lem:Composition3}. Step (c) is the key ingredient of our argument, and we prove it separately.  

Assuming that step (c) holds, we may now conclude the following: The hypothesis~\eqref{Eqn:Lem:Composition} is true for the empty set and all singletons $\{\ell\} \subset [1,L]$ by~\eqref{Eqn:Proof:Lem:Composition1} and~\eqref{Eqn:Proof:Lem:Composition2} respectively. The inductive step~\eqref{Eqn:Proof:Lem:Composition3} holds for all $j \in \set^\text{c}$, and hence the hypothesis~\eqref{Eqn:Lem:Composition} is true for any subset $\set$ with any cardinality $|\set| \in [2,L]$. The next lemma completes the proof by verifying step (c).

\begin{lemma}\label{Lem:InductiveProofLemma}
Let $\set \subset [1,L]$ and $j \in \set^\text{c}$ be arbitrary. Then, 
\begin{equation}\label{Eqn:Lem:InductiveProofLemma}
H(W_j|W_{\set^\text{c} \backslash \{j\}}) = \sum_{\setK \subseteq \set} I_{\setK \cup \{j\}}.
\end{equation}
\end{lemma}
\begin{IEEEproof}
See Appendix~\ref{Sec:Proof:Lem:InductiveProofLemma}.
\end{IEEEproof}

\ifx\doublecolumn\undefined
\begin{figure}[t]
\centering
\includegraphics[width=0.95\textwidth]{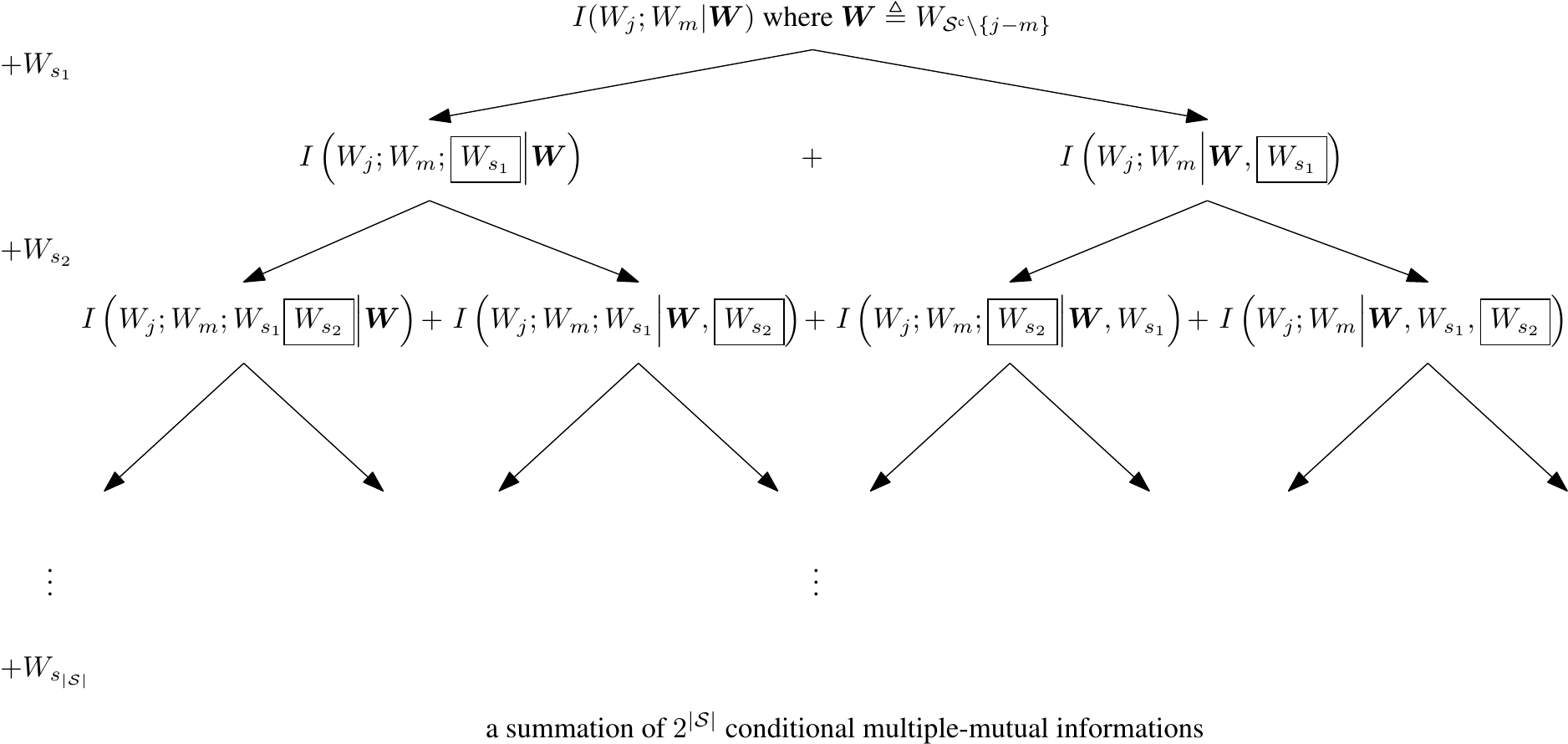}
\caption{Recursive method of including each element of $W_\set =(W_{s_1}, W_{s_2}, \dotsc, W_{s_{|\mathcal{S}|}})$ into $I(W_j;W_m|\boldsymbol{W})$}
\label{fig:expand-mutual-information}
\end{figure}
\else
\begin{figure*}[t]
\centering
\includegraphics[width=0.95\textwidth]{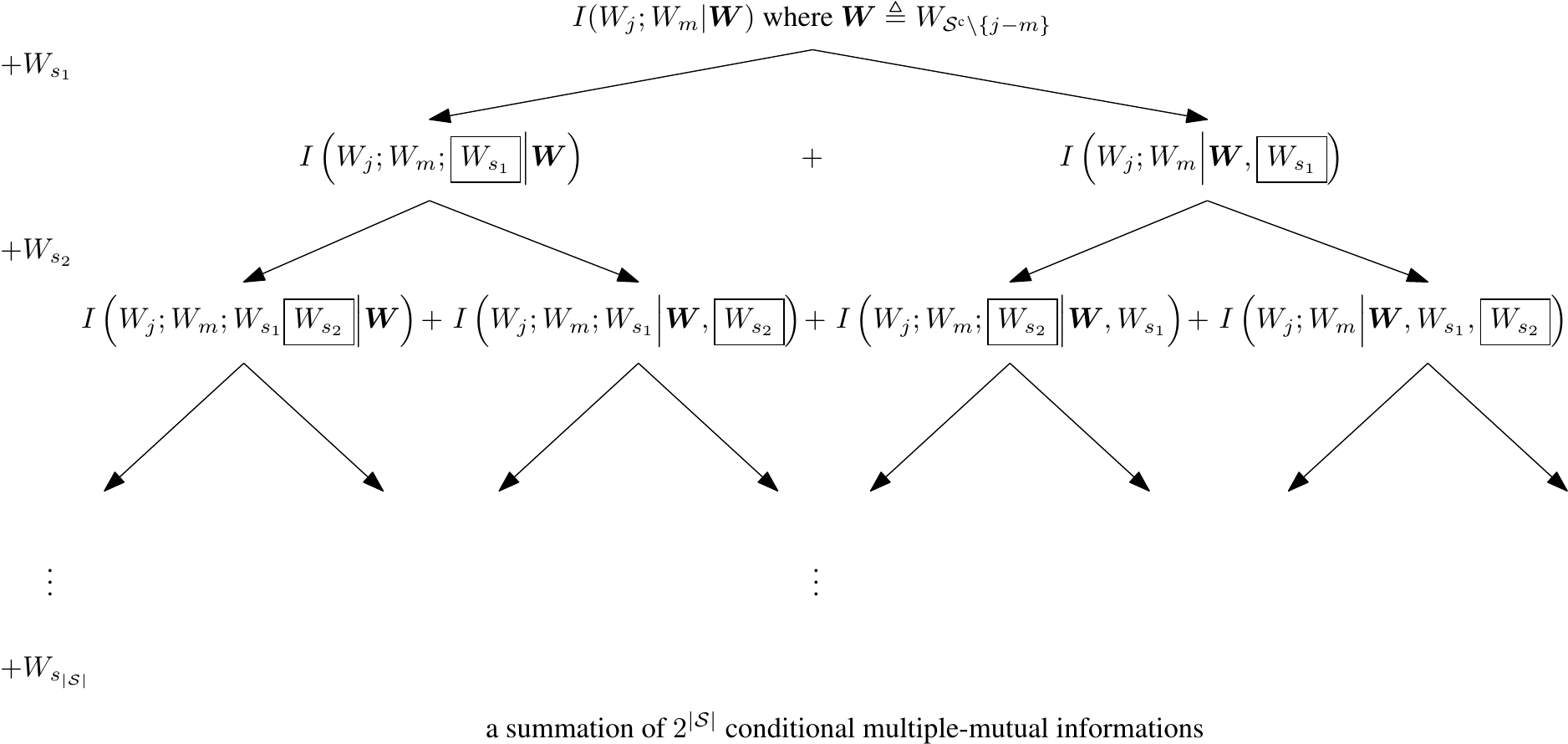}
\caption{Recursive method of including each element of $W_\set =(W_{s_1}, W_{s_2}, \dotsc, W_{s_{|\mathcal{S}|}})$ into $I(W_j;W_m|\boldsymbol{W})$}
\label{fig:expand-mutual-information}
\end{figure*}
\fi


\section{Proof of Lemma~\ref{Lem:InductiveProofLemma}}\label{Sec:Proof:Lem:InductiveProofLemma}

\subsection{A preliminary lemma}

\begin{lemma}\label{proposition:expanding-2-MI}
Let $\set \subset [1,L]$ with $|\set| \leq L-2$ and $j,m \in \setC$ with $j \neq m$ be arbitrary. We have
\begin{equation}
  I(W_j;W_m|W_{\setC \backslash \{j,m\}}) = \sum_{\setK \subseteq \set} I_{\setK \cup \{j,m\}}.\label{eq:summation}
\end{equation}
\end{lemma}

\begin{IEEEproof}
We first recall a useful identity by Hekstra and Willems~\cite[Eqn.~(10b)]{hekstrawillems89}: 
\ifx\doublecolumn\undefined
\begin{align}
I(W_{\ell_1}; W_{\ell_2} ; \dotsm ; W_{\ell_k}|W_{\mathcal{T}}) &= 
I(W_{\ell_1};W_{\ell_2}; \dotsm ; W_{\ell_k};W_{\ell_{k+1}} | W_{\mathcal{T}}) \notag\\
&\quad +
I( W_{\ell_1}; W_{\ell_2}; \dotsm ;W_{\ell_k} | W_{\mathcal{T}},W_{\ell_{k+1}}),\label{eq:expand-mutual-information}
\end{align}
\else
\begin{align}
\notag
I(W_{\ell_1};& W_{\ell_2} ; \dotsm ; W_{\ell_k}|W_{\mathcal{T}}) \\
\notag
&= 
I(W_{\ell_1};W_{\ell_2}; \dotsm ; W_{\ell_k};W_{\ell_{k+1}} | W_{\mathcal{T}})\\
\label{eq:expand-mutual-information}
&\hspace{6mm}
+
I( W_{\ell_1}; W_{\ell_2}; \dotsm ;W_{\ell_k} | W_{\mathcal{T}},W_{\ell_{k+1}}),
\end{align}
\fi
where $\setK = \{\ell_1,\ell_2,\ldots,\ell_{k+1}\} \subset [1,L]$ and $\mathcal{T} \subseteq [1,L] \backslash \setK$ are arbitrary. 


Consider any subset $\set = \{ s_1, s_2, \dotsc, s_{|\set|}\} \subset [1,L]$ with $|\set| \leq L - 2$, 
and some $j,m \in \setC$ with $j \neq m$. We now take $I(W_j;W_m|W_{\setC \backslash \{j,m\}})$ and recursively include each element of $\set$, starting with $s_1$, using Hekstra and Willems' identity~\eqref{eq:expand-mutual-information}.  The procedure is depicted in Figure~\ref{fig:expand-mutual-information}, and it concludes with an expansion consisting of $2^{|\set|}$ conditional-multiple-mutual-information terms:
\ifx\doublecolumn\undefined
\begin{equation}
I(W_j;W_m|W_{\setC \backslash \{j,m\}})
=
\sum_{\setK \subseteq \set} 
I(W_j;W_m;W_{k_1}; \dotsm ; W_{k_{|\setK|}} | W_{\setC \backslash \{j,m
\}},W_{\set \setminus \setK}), \label{Eqn:expanding-2-MI-1}
\end{equation}
\else
\begin{align}
\notag
& I(W_j;W_m|W_{\setC \backslash \{j,m\}})\\
\label{Eqn:expanding-2-MI-1}
&=
\sum_{\setK \subseteq \set} 
I(W_j;W_m;W_{k_1}; \dotsm ; W_{k_{|\setK|}} | W_{\setC \backslash \{j,m
\}},W_{\set \setminus \setK}),
\end{align}
\fi
where the sum on the right hand side is taken over all subsets of the form $\setK = \{k_1,k_2, \dotsc, k_{|\setK|}\} \subseteq \set$. Since $\setK \subseteq \set$ and $j,m \in \setC$, we have 
\begin{align*}
(\setC \backslash \{j,m\}) \cup (\set \backslash \setK )
&= 
((\setC \backslash \setK) \cup (\set \backslash \setK)) \backslash \{j,m\}\\
&= 
[1,L] \backslash ( \setK \cup \{j,m\} \big),
\end{align*}
and~\eqref{Eqn:expanding-2-MI-1} simplifies to 
\ifx\doublecolumn\undefined
\begin{align*}
 I(W_j;W_m|W_{\setC \backslash \{j,m\}})
%
%
&= 
\sum_{\setK \subseteq \set} 
I(W_j;W_m;W_{k_1}; \dotsm ; W_{k_{|\setK|}} | W_{[1:L] \setminus (\setK \cup \{j,m\})})\\
%
%
&=
\sum_{\setK \subseteq \set} I_{\setK \cup \{j,m\}}. \tag*{\IEEEQEDhere\ }
\end{align*}
\else
\begin{align*}
\notag
& I(W_j;W_m|W_{\setC \backslash \{j,m\}})\\
%
%
&= 
\sum_{\setK \subseteq \set} 
I(W_j;W_m;W_{k_1}; \dotsm ; W_{k_{|\setK|}} | W_{[1:L] \setminus (\setK \cup \{j,m\})})\\
%
%
&=
\sum_{\setK \subseteq \set} I_{\setK \cup \{j,m\}}. \tag*{\IEEEQEDhere\ }
\end{align*}
\fi

\end{IEEEproof}


\subsection{Proof of Lemma~\ref{Lem:InductiveProofLemma}}

As before, we use induction to prove~\eqref{Eqn:Lem:InductiveProofLemma}: We first show that~\eqref{Eqn:Lem:InductiveProofLemma} holds for all subsets $\set$ with cardinality $0$ and $1$. We then show that the truth of~\eqref{Eqn:Lem:InductiveProofLemma} for any subset $\set \subset [1,L]$ implies the truth of~\eqref{Eqn:Lem:InductiveProofLemma} for all subsets $\set' \subseteq [1,L]$ of cardinality $|\set'| = |\set| + 1$. 

Starting with the empty set, $\set = \emptyset$, we have
\begin{equation}
H(W_j|W_{\setC\backslash \{j\}}) = H(W_j | W_{\{j\}^\text{c}}) = \sum_{\setK \subseteq \emptyset} I_{\setK \cup \{j\}} = I_{\{j\}}.
\end{equation}
Now consider any $\ell \in [1,L]$ and $\set = \{\ell\}$. We have
\begin{align*}
H(W_j|W_{\setC \backslash \{j\}})
&= 
H(W_j|W_{[1,L] \backslash \{\ell,j\}})\\
&= 
H(W_j|W_{\{j\}^\text{c}})
+ I(W_j ; W_\ell | W_{[1,L] \backslash \{\ell,j\}})\\
&= 
I_{\{j\}} + I_{\{\ell,j\}}\\
&= \sum_{\setK \subseteq \{\ell\}} I_{\setK \cup \{j\}}.
\end{align*}

Suppose now that we are given $\set \subset [1,L]$ with $|\set| \leq L - 2$ such that~\eqref{Eqn:Lem:InductiveProofLemma} holds, i.e.,  
\begin{equation}\label{Eqn:Proof:Lem:InductiveProofLemma-1}
H(W_j | W_{\setC \backslash \{j\}}) 
=
\sum_{\setK \subseteq \set} I_{\setK \cup \{j\}}.
\end{equation}
Pick any $\ell \in \setC$ with $\ell \neq j$. We now prove~\eqref{Eqn:Lem:InductiveProofLemma} for the set $\set \cup \{\ell\}$. We have
\ifx\doublecolumn\undefined
\begin{align}
\notag
H(W_j | W_{(\set\cup \{\ell\})^\text{c} \backslash \{j\}}) &=
H(W_j|W_{\setC\backslash \{\ell,j\}})\\
\notag
&\step{a}{=} 
H(W_j|W_{\setC\backslash\{j\}}) 
+ I(W_j;W_\ell|W_{\setC \backslash \{\ell,j\}}) \\
\notag
&\step{b}{=} 
\sum_{\setK \subseteq \set} I_{\setK \cup \{j\}}
+ \sum_{\setK \subseteq \set} I_{\setK \cup \{\ell,j\}} \\
\label{Eqn:Proof:Lem:InductiveProofLemma-2}
&= 
\sum_{\setK \subseteq \set \cup \{\ell\}} 
I_{\setK\cup \{j\}},
\end{align}
\else
\begin{align}
\notag
H(W_j | W_{(\set\cup \{\ell\})^\text{c} \backslash \{j\}}) &=
H(W_j|W_{\setC\backslash \{\ell,j\}})\\
\notag
&\step{a}{=} 
H(W_j|W_{\setC\backslash\{j\}})\\ 
\notag
&\quad + I(W_j;W_\ell|W_{\setC \backslash \{\ell,j\}}) \\
\notag
&\step{b}{=} 
\sum_{\setK \subseteq \set} I_{\setK \cup \{j\}}
+ \sum_{\setK \subseteq \set} I_{\setK \cup \{\ell,j\}} \\
\label{Eqn:Proof:Lem:InductiveProofLemma-2}
&= 
\sum_{\setK \subseteq \set \cup \{\ell\}} 
I_{\setK\cup \{j\}},
\end{align}
\fi
where (a) uses the chain rule for entropy and (b) applies~\eqref{Eqn:Proof:Lem:InductiveProofLemma-1} and Lemma~\ref{proposition:expanding-2-MI} and $j \in \setC$.

We may now conclude the following from the above argument: The hypothesis~\eqref{Eqn:Lem:InductiveProofLemma} is true for the empty set and all singletons $\set = \{\ell\}$. The inductive step~\eqref{Eqn:Proof:Lem:InductiveProofLemma-2} holds for any $\ell \in \setC \backslash \{j\}$ and, therefore, the hypothesis~\eqref{Eqn:Lem:InductiveProofLemma} is true for any set with any cardinality $|\set| \in [2,L-1]$. \hfill $\blacksquare $


\section{Proof of Theorem~\ref{Thm:Balanced}}\label{App:Proof:Thm:Balanced}

If $L = 2$, then $\Ps = \mathcal{P}$ and the theorem is trivial. Suppose that $L \geq 3$ and $(W_1,\ldots,W_L) \sim p \in \mathcal{P}_\text{bal}$. By Lemma~\ref{Lem:Ps-2}, we need only prove that $\bm{r}^*(p) \in \R(p)$. We start the proof with a useful lemma that represents the rate tuple $\bm{r}^*(p)$ as a weighted sum of conditional multiple-mutual informations. 

Let $\bm{r}^\dag(p) = (r^\dag_1,\ldots,r^\dag_L)$ be defined by
\begin{equation}\label{Eqn:Proof:Lem:Rdagell}
r^\dag_\ell := \sum_{\setK \subset [1,L]} J_\ell(\setK),
\quad \ell \in [1,L],
\end{equation}
where
\begin{equation}\label{Eqn:Proof:Thm:Balanced:JK}
J_\ell(\setK) := 
\left\{
\def\arraystretch{2.4}
\begin{array}{ll}
\left(\dfrac{L- |\setK|}{L-1}\right) I_\setK, &\text{if } \ell \in \setK\\
\left(\dfrac{1-|\setK|}{L-1}\right) I_\setK, &\text{otherwise.}
\end{array}
\right.
\end{equation}

\begin{lemma}\label{Lem:Rdag} $\bm{r}^\dag(p) = \bm{r}^*(p)$ for all $p \in \mathcal{P}$.
\end{lemma}
\begin{IEEEproof}
See Appendix~\ref{Sec:Proof:Lem:Rdag}.
\end{IEEEproof}

We now show that $\bm{r}^\dag(p) \in \R(p)$ by arguing that $\bm{r}^\dag(p)$ satisfies all of the inequalities in~\eqref{Eqn:R}---the inequalities defining $\R(p)$---whenever $p \in \mathcal{P}_\text{bal}$. We first notice that Lemma~\ref{Lem:Rdag} implies that 
\begin{equation*}
\sum_{i \in \ellc} r^\dag_i = H(W_\ellc|W_\ell),
\quad 
\forall\ \ell \in [1,L].
\end{equation*}
Thus, we need only check the inequality in~\eqref{Eqn:R} for all $\set \subset [1,L]$ with cardinality $|\set| \leq L-2$. 

Let $\set \subset [1,L]$ be arbitrary subset with $|\set| \leq L-2$. Consider the sum
\begin{align}
\sum_{i \in \set} r^\dag_i 
\notag
&\step{a}{=}
\sum_{i \in \set}
\sum_{\setK \subset[1,L]} J_i(\setK)\\
\label{Eqn:Proof:Thm:Balanced:1}
&\step{b}{=}
\sum_{k = 1}^{L-1} 
\Gamma_k.
\end{align}
Step (a) follows from~\eqref{Eqn:Proof:Lem:Rdagell}, and in step (b) we define 
\begin{equation*}
\Gamma_k := 
\sum_{\substack{\setK \subset[1,L]\\\text{s.t. }|\setK|=k}} 
\sum_{i \in \set}
J_i(\setK).
\end{equation*}
The next lemma invokes the balanced source assumption and is a key step in the proof.

\begin{lemma}\label{Lem:Gammak}
Fix $\set \subset [1,L]$ with $|\set| \leq L-2$. If $p \in \mathcal{P}_\text{bal}$, then 
\begin{equation*}
\Gamma_k
\geq
\left\{
\begin{array}{cc}
\sum\limits_{\substack{\setK \subseteq \set\\ \text{s.t.} |\setK| = k}} I_\setK,
& 1 \leq k \leq |\set|,\\
%
%
0, & |\set| < k \leq L-1.
\end{array}
\right.
\end{equation*}
\end{lemma}
\begin{IEEEproof}
See Appendix~\ref{Sec:Proof:Lem:Gammak}.
\end{IEEEproof}

Continuing on from~\eqref{Eqn:Proof:Thm:Balanced:1}, we have
\begin{align}
\sum_{i \in \set} r^\dag_i 
\notag
=
\sum_{k=1}^{L-1} \Gamma_k
\notag
\step{a}{\geq}
\sum_{k=1}^{|\set|}
\sum_{\substack{\setK \subseteq \set\\ \text{s.t.} |\setK| = k}} 
I_\setK
\step{b}{=} 
H(W_\set|W_{\set^\text{c}}),
\end{align}
where (a) applies Lemma~\ref{Lem:Gammak} and (b) applies Lemma~\ref{Lem:Composition}. 
\hfill $\blacksquare$


\section{Proof of Lemma~\ref{Lem:Rdag}}\label{Sec:Proof:Lem:Rdag}

Recall that $\bm{r}^*(p)$, defined in~\eqref{Eqn:Rstar}, is the unique solution to\begin{equation}\label{Eqn:Proof:Lem:Rdag1}
\sum_{i \in \ellc} r^*_i = H(W_\ellc|W_\ell),
\quad \forall\ \ell\ \in [1,L]. 
\end{equation}
Now fix $\ell$ and consider the same sum over $i \in \ellc$, but with $\bm{r}^*(p)$ replaced by $\bm{r}^\dag(p)$. We have
\begin{align}
\sum_{i \in \ellc} r^\dag_i 
\notag
&\step{a}{=} 
\sum_{i \in \ellc} \sum_{\setK \subset [1,L]} J_i(\setK)\\
\notag
&\step{b}{=} 
\sum_{\substack{\setK \subset [1,L]\\ \setK \ni \ell}} 
\sum_{i \in \ellc} J_i(\setK)
+
\sum_{\substack{\setK \subset [1,L]\\ \setK \not\owns \ell}} 
\sum_{i \in \ellc}  J_i(\setK)\\
\notag
&\step{c}{=} 
\sum_{\setK \subseteq \ellc} I_\setK\\
\label{Eqn:Proof:Lem:Rdag2}
&\step{d}{=} H(W_\ellc |W_\ell).
\end{align}
Lemma~\eqref{Lem:Rdag} now follows directly from~\eqref{Eqn:Proof:Lem:Rdag2} and the uniqueness of $\bm{r}^*$. Notes:
\begin{enumerate}
\item[a.] Substitute $r^\dag_i$ from~\eqref{Eqn:Proof:Lem:Rdagell}.
\item[b.] Split the summation over the strict subsets $\setK \subset [1,L]$ into two groups: those subsets $\setK$ that own $\ell$, and those $\setK$ that do not own $\ell$. 
\item[c.] Consider the sum over subsets $\setK$ that do not own $\ell$ (the second pair of sums in step (b)): The inner sum over $i$ includes $|\setK|$ elements with $i \in \setK$ and 
\begin{equation*}
J_i(\setK) = \left(\frac{L-|K|}{L-1}\right) I_\setK.
\end{equation*}
The remaining $(L-1-|\setK|)$ elements with $i \notin \setK$ have
\begin{equation*}
J_i(\setK) = \left(\frac{1-|K|}{L-1}\right) I_\setK.
\end{equation*}

\ifx\doublecolumn\undefined
This observation leads to 
\begin{align*} 
\sum_{\substack{\setK \subset [1,L]\\ \setK \not\owns \ell}} \sum_{i \in \ellc}  J_i(\setK)
&= 
\sum_{\substack{\setK \subset [1,L]\\ \setK \not\owns \ell}}
\left(|\setK| \left(\frac{L-|\setK|}{L-1}\right) I_\setK
+ (L-1-|K|)\left(\frac{1 - |\setK|}{L-1}\right) I_\setK
\right)\\
&= \sum_{\substack{\setK \subset [1,L]\\ \setK \not\owns \ell}} I_\setK.
\end{align*}

\else
This observation leads to the expansion shown in~\eqref{Eqn:Proof:Lem:Rdag:ellCase1}, which, in turn, simplifies to 
\begin{equation*}
\sum_{\substack{\setK \subset [1,L]\\ \setK \not\owns \ell}}
\sum_{i \in \ellc}  J_i(\setK)
= 
\sum_{\substack{\setK \subset [1,L]\\ \setK \not\owns \ell}} I_\setK.
\end{equation*}

\begin{floatEq}
\begin{equation}\label{Eqn:Proof:Lem:Rdag:ellCase1}
\sum_{\substack{\setK \subset [1,L]\\ \setK \not\owns \ell}} \sum_{i \in \ellc}  J_i(\setK)
= 
\sum_{\substack{\setK \subset [1,L]\\ \setK \not\owns \ell}}
\left(|\setK| \left(\frac{L-|\setK|}{L-1}\right) I_\setK
+ (L-1-|K|)\left(\frac{1 - |\setK|}{L-1}\right) I_\setK
\right)
\end{equation}
\end{floatEq}
\fi

Consider the sum over subsets $\setK$ that own $\ell$.  The inner sum over $i$ includes $|\setK|-1$ elements with $i \in \setK$ and $(L-1-|\setK|)$ elements with $i \notin \setK$. In this case, we have
\begin{equation*}
\sum_{\substack{\setK \subset [1,L]\\ \setK \owns \ell}} 
\sum_{i \in \ellc}  J_i(\setK)
= 0.
\end{equation*}

\item[d.] Apply Lemma~\ref{Lem:Composition}.
\end{enumerate}


\section{Proof of Lemma~\ref{Lem:Gammak}}\label{Sec:Proof:Lem:Gammak}

Let $\set = \{s_1,s_2,\ldots,s_{|\set|}\} \subset [1,L]$ be any subset with cardinality $|\set| \leq L - 2$, and let $k \in [1,L-1]$ be arbitrary. Table~\ref{Fig:JKTable} will be a useful visual aid throughout the proof. The table consists of 
\begin{equation*}
\binom{L}{k} := \frac{L!}{(L-k)!k!}
\end{equation*}
rows and $|\set|$ columns---one row for each subset $\setK \subset [1,L]$ with cardinality $k$ and one column for each element of $\set$.
Let $\setK_1,\setK_2,\ldots,\setK_{\binom{L}{k}}$ be any ordering (for example, lexicographic) of all the subsets $\setK$ with cardinality $k$, and let $\setK_i$ be the label for the $i$-th row of the table. Assign to the cell $(\setK_i,s_\ell)$ the value $J_{s_\ell}(\setK_i)$. 

\ifx\doublecolumn\undefined
\begin{table}[t]
\centering
\includegraphics[width=0.55\textwidth]{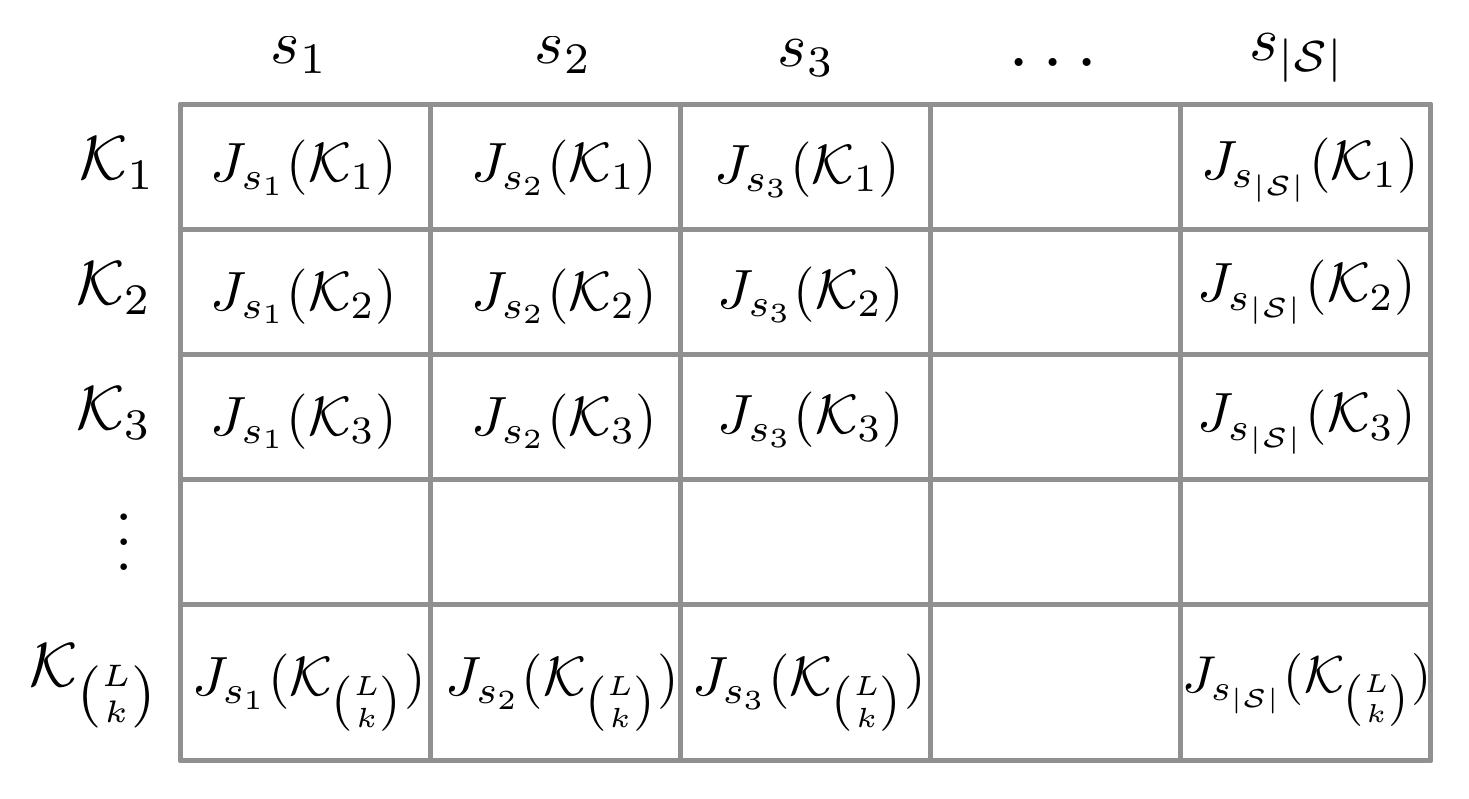}
\caption{Visual aid for the proof of Lemma~{\ref{Lem:Gammak}}.}
\label{Fig:JKTable}
\vspace{-10mm}
\end{table}
\else
\begin{table}[t]
\centering
\includegraphics[width=0.45\textwidth]{03-table1}
\caption{Visual aid for the proof of Lemma~{\ref{Lem:Gammak}}.}
\label{Fig:JKTable}
\vspace{-5mm}
\end{table}
\fi

We may rewrite $\Gamma_k$ as a sum over all cells in Table~\ref{Fig:JKTable},
\begin{equation}\label{Eqn:Proof:Gammak:2}
\Gamma_k 
= 
\sum_{i = 1}^{\binom{L}{k}} 
\sum_{\ell = 1}^{|\set|}
J_{s_\ell}(\setK_i).
\end{equation}
Recall that 
\begin{equation*}
J_{s_\ell}(\setK_i) =
\left\{
\begin{array}{ll}
\ \ \Big(\frac{L-k}{L-1}\Big) I_{\setK_i}, & \text{ if } s_\ell \in \setK_i\\
- \Big(\frac{k-1}{L-1}\Big) I_{\setK_i}, & \text{ if } s_\ell \notin \setK_i.
\end{array}
\right.
\end{equation*} 

\begin{definition}
We call a cell $(\setK_i,s_\ell)$ \emph{positive} if $s_\ell \in \setK_i$ and \emph{negative} otherwise.\footnote{The terms \emph{positive} and \emph{negative} refer to the sign of the coefficient fraction in $J_{s_\ell}(\setK_i)$, and not to the sign of $I_{\setK_i}$.}
\end{definition}

\begin{lemma}\label{Lem:Thm:Ps:1}
In each and every column in Table~\ref{Fig:JKTable}, there are 
\begin{equation*}
\binom{L-1}{k-1}
\quad 
\text{and}
\quad
\binom{L-1}{k}
\end{equation*}
positive and negative cells respectively. 
\end{lemma}

\begin{IEEEproof}
Consider an arbitrary column $s_\ell \in \set$. Recall that there are $\binom{L-1}{k-1}$ ways of selecting $(k-1)$ unordered elements from the set $[1,L]\backslash \{s_\ell\}$. The union of each such selection with $\{s_\ell\}$ forms a subset $\setK_i$ such that $|\setK| = k$ and $\setK_i \owns s_\ell$, so it follows that the column has $\binom{L-1}{k-1}$ positive cells. The remaining 
\begin{equation*}
\binom{L}{k} - \binom{L-1}{k-1} = \binom{L-1}{k}
\end{equation*}
cells in the column are negative.  
\end{IEEEproof}

\begin{lemma}\label{Lem:NumberOfPosAndNegInTable}
Throughout the entire table, there are 
\begin{equation*}
|\set| \binom{L-1}{k-1}
\quad 
\text{and}
\quad
|\set| \binom{L-1}{k}
\end{equation*}
positive and negative cells respectively.
\end{lemma}
\begin{IEEEproof}
The table has $|\set|$ columns and Lemma~\ref{Lem:Thm:Ps:1} holds for every column. 
\end{IEEEproof}

We now prove the Lemma~\ref{Lem:Gammak} individually for each of the following three cases: $k = 1$; $2 \leq k \leq |\set|$; and $|\set| + 1 \leq k \leq L-1$.

\subsection{Case: $k = 1$} 
We trivially have
\begin{equation*}
J_i(\setK) = 
\left\{
\begin{array}{ll}
I_\setK, & \text{ if } \setK = \{i\},\\
0, & \text{ otherwise,}
\end{array}
\right.
\end{equation*}
and therefore
\begin{equation}\label{Eqn:Proof:Lem:Gammak:k2:Roy1a}
\Gamma_1 
= 
\sum_{i \in \set} I_{\{i\}}
=
\sum_{\substack{\setK \subseteq \set\\|\setK| = 1}} I_\setK.
\end{equation}


\subsection{Case: $2 \leq k \leq |\set|$}

\ifx\doublecolumn\undefined
We now show that $\Gamma_k$ is lower bounded as follows:
\begin{align}
\Gamma_k 
\notag
&= 
\sum_{i=1}^{\binom{L}{k}}
\sum_{\ell = 1}^{|\set|}  J_{s_\ell}(\setK_i)\\
\notag
&\step{a}{=}
\sum_{i=1}^{\binom{L}{k}}
\sum_{\ell = 1}^{|\set|} 
\indicator{\setK_i \subseteq \set}\
 J_{s_\ell}(\setK_i) 
+
\sum_{i=1}^{\binom{L}{k}}
\sum_{\ell = 1}^{|\set|}
\big(1-\indicator{\setK_i \subseteq \set}\big)\
J_{s_\ell}(\setK_i) 
\\
\notag
&\step{b}{=}
\sum_{i=1}^{\binom{L}{k}}
\indicator{\setK_i \subseteq \set}\
I_{\setK_i}
+ 
\sum_{i=1}^{\binom{L}{k}}
\indicator{\setK_i \subseteq \set}
\left(\frac{(L-|\set|-1)(k-1)}{L-1}\right) I_{\setK_i}\\
\notag
&\quad
+
\underbrace{\sum_{i=1}^{\binom{L}{k}}
\sum_{\ell = 1}^{|\set|}
\big(1-\indicator{\setK_i \subseteq \set}\big)\
J_{s_\ell}(\setK_i)}_\text{inactive rows}
\\
\notag
&\step{c}{\geq}
\sum_{i=1}^{\binom{L}{k}}
\indicator{\setK_i \subseteq \set}\
I_{\setK_i}
+ 
\binom{|\set|}{k}
\left(\frac{(L-|\set|-1)(k-1)}{L-1}\right) \underline{\mu}_k\\
\notag
&\quad +
\underbrace{\left( |\set| \binom{L-1}{k-1} - k \binom{|\set|}{k} \right) 
\left( \frac{L-k}{L-1} \right) \underline{\mu}_k
- \left( |\set| \binom{L-1}{k} - (|\set| - k) \binom{|\set|}{k} \right) 
\left(\frac{k-1}{L-1}\right) \overline{\mu}_k}_\text{inactive rows}\\
\label{Eqn:Proof:Lem:Gammak:k2}
&\step{d}{=} 
\sum_{i=1}^{\binom{L}{k}} 
\indicator{\setK_i \subseteq \set}\ I_{\setK_i} 
+\left(\frac{k-1}{L-1}\right) \eta.
\end{align}

The next definition and lemma will be useful in explaining the steps leading~\eqref{Eqn:Proof:Lem:Gammak:k2}. 
\else
\begin{floatEq}
\begin{align}
\Gamma_k 
\notag
&= 
\sum_{i=1}^{\binom{L}{k}}
\sum_{\ell = 1}^{|\set|}  J_{s_\ell}(\setK_i)\\
\notag
&\step{a}{=}
\sum_{i=1}^{\binom{L}{k}}
\sum_{\ell = 1}^{|\set|} 
\indicator{\setK_i \subseteq \set}\
 J_{s_\ell}(\setK_i) 
+
\sum_{i=1}^{\binom{L}{k}}
\sum_{\ell = 1}^{|\set|}
\big(1-\indicator{\setK_i \subseteq \set}\big)\
J_{s_\ell}(\setK_i) 
\\
\notag
&\step{b}{=}
\sum_{i=1}^{\binom{L}{k}}
\indicator{\setK_i \subseteq \set}\
I_{\setK_i}
+ 
\sum_{i=1}^{\binom{L}{k}}
\indicator{\setK_i \subseteq \set}
\left(\frac{(L-|\set|-1)(k-1)}{L-1}\right) I_{\setK_i}
+
\underbrace{\sum_{i=1}^{\binom{L}{k}}
\sum_{\ell = 1}^{|\set|}
\big(1-\indicator{\setK_i \subseteq \set}\big)\
J_{s_\ell}(\setK_i)}_\text{inactive rows}
\\
\notag
&\step{c}{\geq}
\sum_{i=1}^{\binom{L}{k}}
\indicator{\setK_i \subseteq \set}\
I_{\setK_i}
+ 
\binom{|\set|}{k}
\left(\frac{(L-|\set|-1)(k-1)}{L-1}\right) \underline{\mu}_k\\
\notag
&\hspace{20mm}
+
\underbrace{\left( |\set| \binom{L-1}{k-1} - k \binom{|\set|}{k} \right) 
\left( \frac{L-k}{L-1} \right) \underline{\mu}_k
- \left( |\set| \binom{L-1}{k} - (|\set| - k) \binom{|\set|}{k} \right) 
\left(\frac{k-1}{L-1}\right) \overline{\mu}_k}_\text{inactive rows}\\
\label{Eqn:Proof:Lem:Gammak:k2}
&\step{d}{=} 
\sum_{i=1}^{\binom{L}{k}} 
\indicator{\setK_i \subseteq \set}\ I_{\setK_i} 
+\left(\frac{k-1}{L-1}\right) \eta
\end{align}
\end{floatEq}

We now show that $\Gamma_k$ is lower bounded as~\eqref{Eqn:Proof:Lem:Gammak:k2}.  The next definition and lemma will be useful in explaining the steps leading~\eqref{Eqn:Proof:Lem:Gammak:k2}. 
\fi

\begin{definition}
We say that row $\setK_i$ of the Table~\ref{Fig:JKTable} is \emph{active} if $\setK_i \subseteq \set$ and \emph{inactive} if $\setK_i \not\subseteq \set$.  
\end{definition}

\begin{lemma}\label{Lem:NumberOfPosAndNegInActiveRows}
In Table~\ref{Fig:JKTable}, there are 
\begin{equation*}
k\binom{|\set|}{k}
\quad 
\text{and}
\quad 
|\set|\binom{L-1}{k-1} - k\binom{|\set|}{k}
\end{equation*}
positive cells in active and inactive rows respectively. Similarly, there are 
\begin{equation*}
(|\set|-k)\binom{|\set|}{k}
\quad 
\text{and}
\quad 
|\set|\binom{L-1}{k} - (|\set|-k)\binom{|\set|}{k}
\end{equation*}
negative cells in active and inactive rows respectively.
\end{lemma}
\begin{IEEEproof}
The are $\binom{|\set|}{k}$ active rows in the table and each active row has $k$ positive cells, so there are $k\binom{|\set|}{k}$ positive cells in active rows. The remaining 
\begin{equation*}
|\set|\binom{L-1}{k-1} - k\binom{|\set|}{k}
\end{equation*}
active cells (here we have used Lemma~\ref{Lem:NumberOfPosAndNegInTable}) belong to inactive rows. Similarly, there are $(|\set| - k) \binom{|\set|}{k}$ negative cells in active rows. The remaining 
\begin{equation*}
|\set|\binom{L-1}{k} - (|\set|-k)\binom{|\set|}{k}
\end{equation*} 
negative cells are in inactive rows. 
\end{IEEEproof}

Notes on~\eqref{Eqn:Proof:Lem:Gammak:k2}:
\begin{itemize}
\item[a.] Split the outer sum (over rows in Table~\ref{Fig:JKTable}) into active and nonactive rows using 
\begin{equation*}
\indicator{\setK_i \subseteq \set} := 
\left\{ 
\begin{array}{ll}
1 & \text{ if } \setK_i \subseteq \set\\
0 & \text{ otherwise.}
\end{array}
\right.
\end{equation*} 
\item[b.] There are $k$ positive cells and $(|\set| - k)$ negative cells in each and every active row. Therefore,  for every $\setK_i \subseteq\set$, 
\begin{align*}
\sum_{\ell=1}^{|\set|} 
 J_{s_\ell}(\setK_i)
&= 
k \left(\frac{L-k}{L-1}\right) I_{\setK_i}
- (|\set|-k)\left(\frac{k-1}{L-1}\right)I_{\setK_i}\\
&= \left(1 + \frac{(L-|\set|-1)(k-1)}{L-1}\right) I_{\setK_i}.
\end{align*}

\item[c.] Use Lemma~\ref{Lem:NumberOfPosAndNegInActiveRows} to count the number of positive and negative cells in the inactive rows (the rightmost pair of sums in step b), and substitute 
\ifx\doublecolumn\undefined
\begin{equation*}
\underline{\mu}_k =\min\{I_{\setK_1},I_{\setK_2},\ldots,I_{\setK_{\binom{L}{k}}}\}\quad
\text{and}\quad
\overline{\mu}_k =\max\{I_{\setK_1},I_{\setK_2},\ldots,I_{\setK_{\binom{L}{k}}}\}.
\end{equation*}
\else
\begin{multline*}
\underline{\mu}_k =\min\{I_{\setK_1},I_{\setK_2},\ldots,I_{\setK_{\binom{L}{k}}}\}\\
\text{and}\quad
\overline{\mu}_k =\max\{I_{\setK_1},I_{\setK_2},\ldots,I_{\setK_{\binom{L}{k}}}\}.
\end{multline*}
\fi

\item[d.] Clean up the terms (outside the sum in step c) into 
\begin{equation*}
\eta := 
\left(\alpha(k,\set) + \frac{1}{k-1} \beta(k,\set)\right) \underline{\mu}_k - \alpha(k,\set)\ \overline{\mu}_k,
\end{equation*}
where
\begin{equation*}
\alpha(k,\set) := |\set| \binom{L-1}{k} - (|\set|-k) \binom{|\set|}{k}
\end{equation*}
and
\begin{equation*}
\beta(k,\set) := |\set| \binom{L-1}{k} - (L-1) \binom{|\set|}{k}.
\end{equation*}
\end{itemize} 

Consider~\eqref{Eqn:Proof:Lem:Gammak:k2}:
\begin{equation*}
\Gamma_k 
\geq 
\left(\frac{k-1}{L-1}\right) \eta
+ \sum_{\substack{\setK \subseteq \set\\\text{s.t. } |\setK| = k}}\hspace{-2mm} I_{\setK},
\end{equation*}
and, in particular, the constants $\alpha(k,\set)$ and $\beta(k,\set)$ that make up $\eta$. We have $\alpha(k,\set) > 0$ and $\beta(k,\set) > 0$, so it follows that $\eta \geq 0$ whenever
\begin{equation}\label{Eqn:Proof:Lem:Gammak:k2:Roy1}
\overline{\mu}_k \leq 
\left(1 + \frac{\beta(k,\set)}{(k-1)\ \alpha(k,\set)}\right) \underline{\mu}_k. 
\end{equation}
The next lemma shows that~\eqref{Eqn:Proof:Lem:Gammak:k2:Roy1} does indeed hold whenever $p \in \mathcal{P}_\text{bal}$, and therefore
\begin{equation}\label{Eqn:Proof:Lem:Gammak:k2:Roy2}
\Gamma_k \geq \sum_{\substack{\setK \subseteq \set\\\text{s.t. } |\setK| = k}} I_\setK.
\end{equation}

\begin{lemma}\label{Lem:RoyBalanced}
Fix $(W_1,\ldots,W_L) \sim p$ with $p \in \mathcal{P}_\text{bal}$ and $2 \leq k \leq L -2$. For any subset $\setT \subset [1,L]$  with $k \leq |\setT| \leq L- 2$, we have \eqref{Eqn:Proof:Lem:Gammak:k2:Roy1}.
\end{lemma}
\begin{IEEEproof}
See Appendix~\ref{Sec:Proof:Lem:RoyBalanced}.
\end{IEEEproof}


\subsection{Case: $|\set| \leq k \leq L-1$}

We have
\begin{align}
\notag
\Gamma_k &=
\sum_{i=1}^{\binom{L}{k}} \sum_{\ell = 1}^{|\set|} J_{s_\ell}(\setK_i)\\
\notag
&\step{a}{\geq} \binom{L-1}{k-1} \left(\frac{L-k}{L-1}\right) \underline{\mu}_k 
- \binom{L-1}{k} \left(\frac{k-1}{L-1}\right)\overline{\mu}_k\\
\notag
&= \frac{(L-2)!}{(L-k-1)!k!} \left(k \underline{\mu}_k - (k-1) \overline{\mu}_k \right)\\
\label{Eqn:Proof:Lem:Gammak:k2:Roy3}
&\step{b}{\geq} 0.
\end{align}

Notes:
\begin{itemize}
\item[a.] Use Lemma~\ref{Lem:NumberOfPosAndNegInTable} to count the number of positive and negative cells Table~\ref{Fig:JKTable}, and bound the corresponding conditional multiple-mutual informations by $\overline{\mu}_k$ and $\underline{\mu}_k$. 
\item[b.] For all $k \in [2,L]$, we have
\begin{equation*}
\underline{\mu}_k 
\step{b.1}{\leq} 
\overline{\mu}_k 
\step{b.2}{\leq} 
\left( 1 + \frac{1}{k} \left( \frac{L-1}{2L-k-3} \right) \right) \underline{\mu}_k
\step{b.3}{\leq} 
\left(\frac{k}{k-1}\right) \underline{\mu}_k.
\end{equation*}
Step (b.1) follows by definition of $\underline{\mu}_k$ and $\overline{\mu}_k$; step (b.2) follows because the source is balanced, $p \in \mathcal{P}_\text{bal}$; and step (b.3) follows because
\begin{equation*}
1 + \frac{1}{k} \left(\frac{L-1}{2L-k-3}\right)
\leq \frac{1}{k-1}, \quad \forall\ k \in [2,L-1].
\end{equation*} 
It follows that $k \underline{\mu}_k - (k-1) \overline{\mu}_k \geq 0$, since $0 \leq \underline{\mu}_k \leq \overline{\mu}_k$.
\end{itemize}


\section{Proof of Lemma~\ref{Lem:RoyBalanced}}\label{Sec:Proof:Lem:RoyBalanced}

Fix $2 \leq k \leq L -2$. Let $\setT \subset [1,L]$ be any subset with cardinality $k \leq |\setT| \leq L - 2$. We have
\begin{equation*}
\frac{\beta(k,\setT)}{\alpha(k,\setT)}
\geq
\frac{k-1}{k} \left(\frac{L-1}{2L - k - 3}\right),
\end{equation*}
and it then follows that $p \in \mathcal{P}_\text{bal}$ implies
\begin{equation*}
\overline{\mu}_k \leq \left( 1 + \frac{\beta(k,\setT)}{(k-1) \alpha(k,\setT)} \right) \underline{\mu}_k. \tag*{$\blacksquare$}
\end{equation*}


\section{Proof of Theorem~\ref{Thm:Storage}}\label{Sec:Proof:Thm:Storage}

The centralised storage problem is equivalent to the distributed source coding problem in Appendix~\ref{Sec:SourceCoding}. By Lemma~\ref{Lem:Rdag}, a total storage rate $r_\Sigma \geq 0$ is achievable if and only if there exists a rate tuple $\bm{r} \in \text{int}(\R(p))$ such that $r_\Sigma \geq \|\bm{r}\|$. The optimal total storage rate $r_\Sigma^*$ is then
\begin{equation*}
r_\Sigma^* = \inf_{\bm{r} \in \text{int}(\R(p))}  \|\bm{r}\| = \min_{\bm{r} \in \R(p)}  \|\bm{r}\|. \tag*{$\blacksquare$}
\end{equation*}



\end{document}